\documentclass[conference]{IEEEtran}

\usepackage[noend]{algpseudocode}
\usepackage{microtype}
\usepackage{bbm}
\usepackage{graphicx}
\usepackage[caption=false,font=footnotesize, captionskip=0pt]{subfig}
\usepackage{hyperref}
\usepackage[compress]{cleveref}
\usepackage{mathrsfs}
\usepackage{bm}
\usepackage{balance}
\usepackage{booktabs}
\usepackage{xcolor}
\usepackage[normalem]{ulem}
\usepackage{floatrow}
\usepackage{cite}

\graphicspath{./figures/}

\crefname{section}{Section}{Sections}
\Crefname{section}{Section}{Sections}
\crefname{figure}{Fig.}{Figs.}
\Crefname{figure}{Fig.}{Figs.}
\crefname{subfigure}{Fig.}{Figs.}
\Crefname{subfigure}{Fig.}{Figs.}
\crefname{table}{Table}{Tables}
\Crefname{table}{Table}{Tables}
\crefrangelabelformat{subfigure}{#3#1#4--#5(\crefstripprefix{#1}{#2}#6}

\newcommand{\todo}[1]{\textcolor{red}{\{\{#1\}\}}}
\newcommand{\brandon}[1]{\textcolor{teal}{[Brandon: #1]}}

\newcommand{\revision}[2]{\textcolor{black}{#2}}
\newcommand{\revisioncolor}[1]{\textcolor{black}{#1}}

\newcommand{\squeeze}[1]{\textls[-10]{#1}}
\newcommand{\squeezemore}[1]{\textls[-20]{#1}}

\newcommand{\Partition}[2]{\ensuremath{\textsc{partition}(#1, #2)}}
\newcommand{\LayoutSpec}{\ensuremath{\mathscr{L}}}

\def\BibTeX{{\rm B\kern-.05em{\sc i\kern-.025em b}\kern-.08em
    T\kern-.1667em\lower.7ex\hbox{E}\kern-.125emX}}

\begin{document}
\bstctlcite{IEEEexample:BSTcontrol}

\title{TASM: A Tile-Based Storage Manager for Video Analytics}

\begin{sloppypar}

\author{
\IEEEauthorblockN{Maureen Daum\IEEEauthorrefmark{1}, Brandon Haynes\IEEEauthorrefmark{2}, Dong He\IEEEauthorrefmark{1}, Amrita Mazumdar\IEEEauthorrefmark{1}, Magdalena Balazinska\IEEEauthorrefmark{1}}
\IEEEauthorblockA{\IEEEauthorrefmark{1}Paul G. Allen School of Computer Science \& Engineering, University of Washington \\
\{mdaum, donghe, amrita, magda\}@cs.washington.edu}
\IEEEauthorblockA{\IEEEauthorrefmark{2}Gray Systems Lab, Microsoft \\
Brandon.Haynes@microsoft.com}
}

% \author{Alvin Cheung}
% \affiliation{
%     \institution{Department of Electrical Engineering and Computer Sciences}
%     \institution{University of California, Berkeley}
% }
% \email{akcheung@cs.berkeley.edu}

% \input{icde_response}
% \newpage

\maketitle

\begin{abstract}

 % The amount of video data being produced is rapidly growing. At the same time, advances in machine learning and computer vision have enabled applications to query over the contents of videos. For example, an ornithology application may retrieve birds of various species from a nature video. However, m

Modern video data management systems store videos as a single encoded file, which significantly limits possible storage level optimizations. We design, implement, and evaluate TASM, a new tile-based storage manager for video data. TASM uses a feature in modern video codecs called ``tiles'' that enables spatial random access into encoded videos. TASM physically tunes stored videos by optimizing their tile layouts given the video content and a query workload. Additionally, TASM dynamically tunes that layout in response to changes in the query workload or if the query workload and video contents are incrementally discovered. Finally, TASM also produces efficient initial tile layouts for newly ingested videos. We demonstrate that TASM can speed up \revision{R3D3}{subframe selection} queries by an average of over 50\% and up to 94\%. TASM can also improve the throughput of the full scan phase of object detection queries by up to 2$\times$.

%  does not provide opportunities to optimize queries for spatial subsets of videos. We propose utilizing a feature in modern video codecs called ``tiles'' that enable spatial random access into encoded videos. We present the design of TASM, a tile-based storage manager, and describe techniques it uses to optimize the physical layout of videos for various query workloads. We demonstrate how TASM can significantly improve the performance of queries over videos when the workload is known, as well as how it can incrementally adapt the physical layout of videos to improve performance even when the workload is not known. Layouts picked by TASM speed up individual queries by an average of 51\% and up to 94\% while maintaining good quality. TASM can also accelerate the preprocessing phase of object detection queries by up to $2\times$ while maintaining accuracy.

\end{abstract}

%!TEX root = paper.tex

% % % NON-EVALUATION FIGURES

\newcommand{\tileLayoutDurationFigure}{
    \renewcommand{\FBbskip}{-1.5em}
    \begin{figure}[t!]
        \centering
        \subfloat[Long layout duration]{\includegraphics[width=0.47\columnwidth]{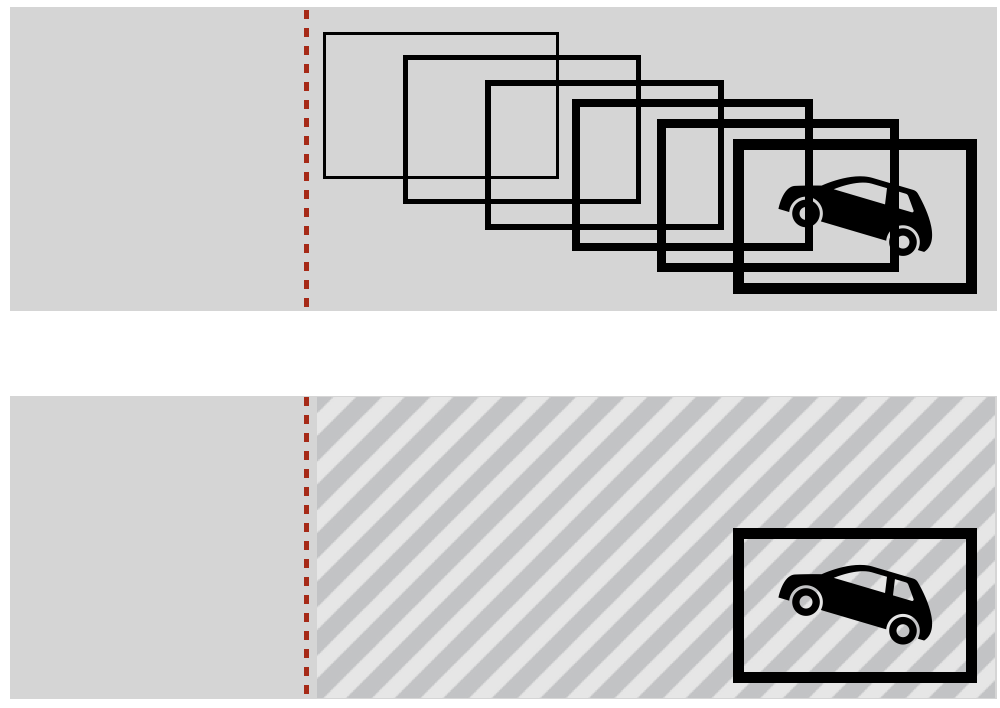}%
            \label{subfig:longLayout}}%
        \hfil
        \subfloat[Short layout duration]{\includegraphics[width=0.47\columnwidth]{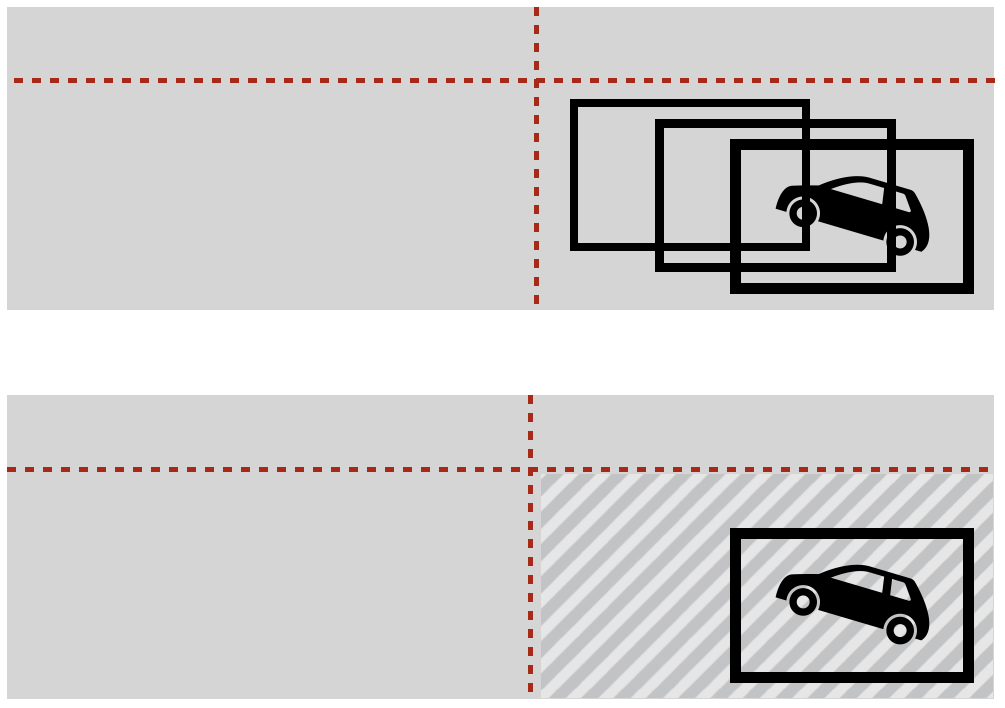}%
            \label{subfig:shortLayout}}%
        \vspace{-.5em}
        \caption[]{\subref{subfig:longLayout} shows how more pixels must be decoded on each individual frame when a tile layout extends for many frames compared to \subref{subfig:shortLayout} where fewer frames have the same layout. The boxes show the location of the car on later frames, and the dashed lines show the tile boundaries. The striped region indicates the tile that would be decoded for a query targeting cars.}
        % \vspace{-1.65em}
        \label{fig:tileLayoutDurationExample}
    \end{figure}
}

\newcommand{\comparingCustomLayoutsFigure}{
    \renewcommand{\FBbskip}{-1.5em}
    \begin{figure}[t!]
        \centering
        \subfloat[Uniform 2x4 layout]{\includegraphics[width=0.45\columnwidth]{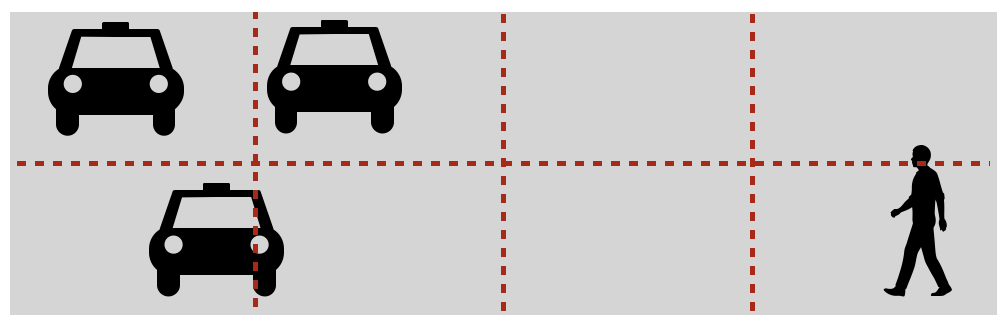}%
            \label{subfig:uniformLayoutCarPed}}%
        \hfil
        \subfloat[Layout~around~cars~\&~people]{\includegraphics[width=0.45\columnwidth]{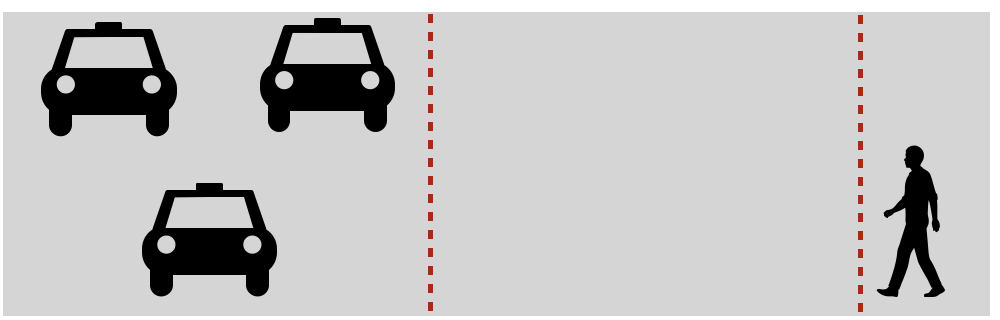}%
            \label{subfig:customLayoutCarPed}}%

        \vspace{-.5em}
        \subfloat[Tile layout around cars]{\includegraphics[width=0.45\columnwidth]{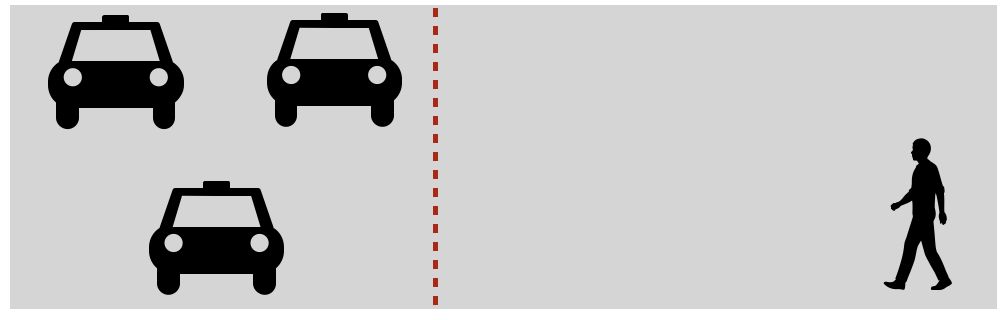}%
            \label{subfig:customLayoutCar}}%
        \hfil
        \subfloat[Tile layout around people]{\includegraphics[width=0.45\columnwidth]{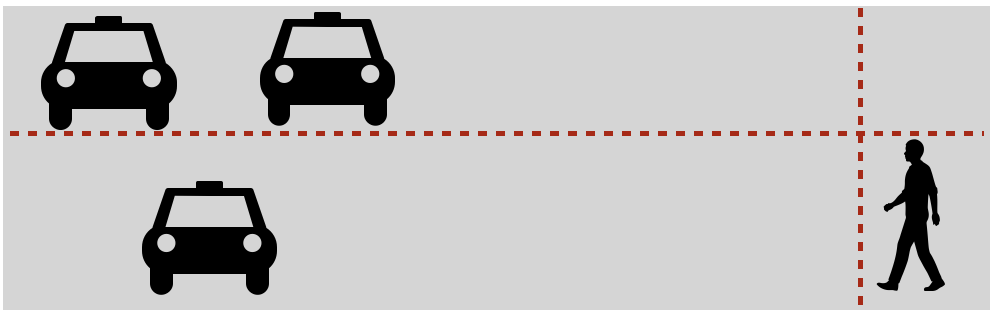}%
            \label{subfig:customLayoutPed}}%
        \vspace{-.5em}
        \caption[]{Various ways to tile a frame. \subref{subfig:uniformLayoutCarPed} is a uniform layout, while \subref{subfig:customLayoutCarPed}-\subref{subfig:customLayoutPed} are non-uniform layouts. Depending on which objects are targeted, different layouts will be more efficient.}
        % \vspace{-1em}
        \label{fig:comparingCustomLayouts}
    \end{figure}
    % Retrieving people from \subref{subfig:uniformLayoutCarPed} requires decoding fewer pixels than from \subref{subfig:customLayoutCarPed} demonstrating how in some cases custom layouts can be less efficient than uniform layouts.
}

% % % EVALUATION FIGURES

\newcommand{\tileAndFilesFigure}{
    \renewcommand{\FBbskip}{-1.5em}
    \begin{figure}[t!]
        \centering
        \subfloat[][]{\includegraphics[width=0.33\textwidth]{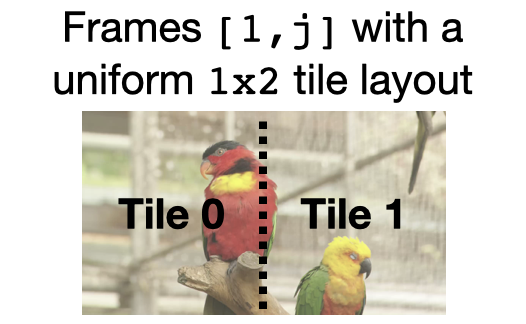}%
            \label{subfig:firstFrames}}%
        \hfil
        \subfloat[][]{\includegraphics[width=0.33\textwidth]{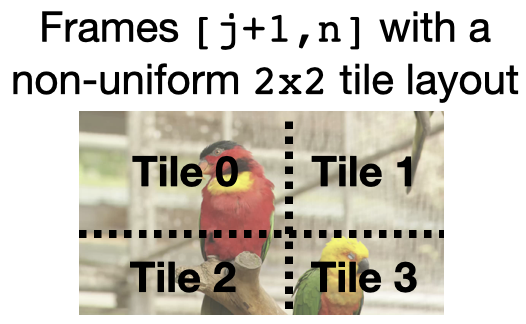}%
            \label{subfig:lastFrames}}%
        \hfil
        \subfloat[][]{\includegraphics[width=0.33\textwidth]{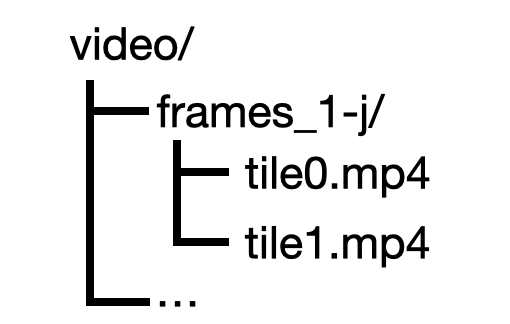}%
            \label{subfig:directoryStructure}}%
        \vspace{-.5em}
        \caption[]{
            % \brandon{Maybe bump the font size above the frames up a bit?  You have the horizontal space and it's a title-page figure.}
            Video partitioned into tiles.
            \subref{subfig:firstFrames} shows the first $j$ frames partitioned with a uniform 1$\times$2 layout.
            \subref{subfig:lastFrames} shows frames partitioned with a non-uniform 2$\times$2 layout.
            \subref{subfig:directoryStructure} shows a directory hierarchy. Video stored at {\footnotesize \texttt{video/frames\_1-j/tile0.mp4}} contains the left half of frames $[1, j]$.}
        % \vspace{-2em}
        \label{fig:tileAndFiles}
    \end{figure}
}

\newcommand{\queryToTilesToDecodeFigure}{
    \begin{figure}[t!]
        \centering
        \includegraphics[width=0.95\columnwidth]{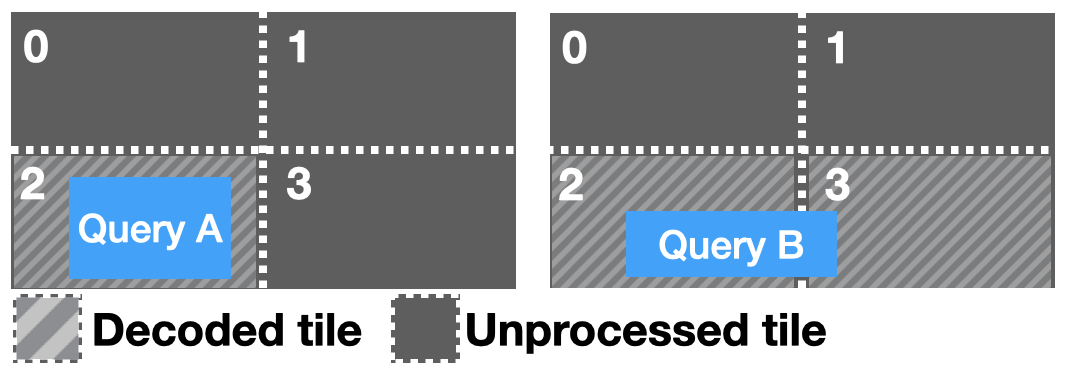}
        \vspace{-1em}
        \caption{Tiles to decode for various queries. Only tile 2 needs to be decoded for query A. Tiles 2 and 3 must be decoded for query B because it requests a region that lies in both tiles.}
        \vspace{-1em}
        \label{fig:queryToTiles}
    \end{figure}
}

\newcommand{\fineVsCoarseGrainFigure}{
    \renewcommand{\FBbskip}{-1.5em}
    \begin{figure}[t!]
        \centering
        \subfloat[Fine-grained tiles]{\includegraphics[width=0.47\columnwidth]{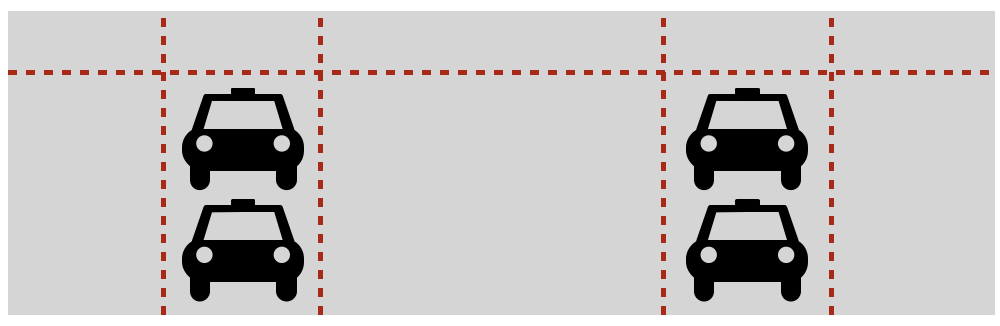}%
            \label{subfig:fineGrainTiles}}%
        \hfil
        \subfloat[Coarse-grained tiles]{\includegraphics[width=0.47\columnwidth]{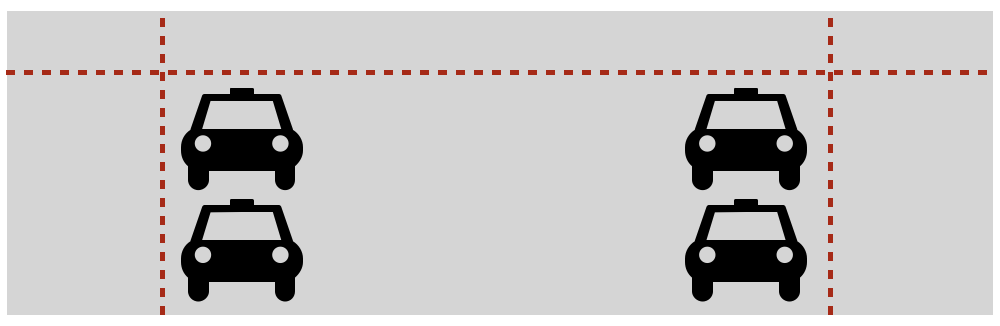}%
            \label{subfig:coarseGrainTiles}}%
        \vspace{-.5em}
        \caption[]{Non-uniform tile layout around cars using \subref{subfig:fineGrainTiles} fine-grained tiles, or \subref{subfig:coarseGrainTiles} coarse-grained tiles.}
        % \vspace{-1em}
        \label{fig:fineVsCoarseTiles}
    \end{figure}
}

\newcommand{\tileGranularityPlots}{
    \renewcommand{\FBbskip}{-1.5em}
    \begin{figure}
        \centering

        \subfloat{\includegraphics[width=0.4\columnwidth]{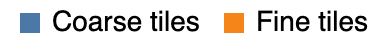}}%
        \vspace{-1em}
        \setcounter{subfigure}{0}

        % {0.3\columnwidth}
        \subfloat[Same]{\includegraphics[height=3cm]{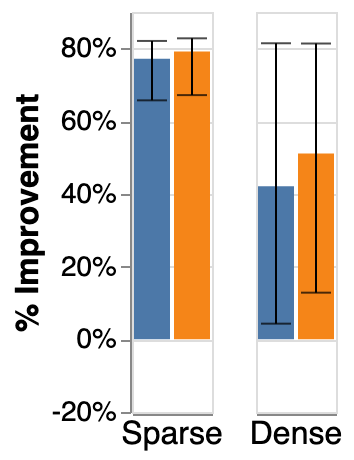}%
            \label{subfig:same}}%
        \hfil
        % {0.21\columnwidth}
        \subfloat[Different]{\includegraphics[height=3cm]{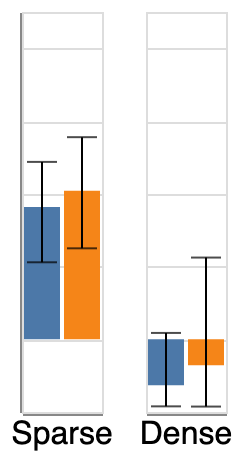}%
            \label{subfig:different}}%
        \hfil
        % {0.21\columnwidth}
        \subfloat[All]{\includegraphics[height=3cm]{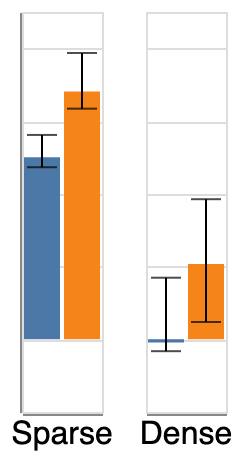}%
            \label{subfig:all}}%
        \hfil
        % {0.21\columnwidth}
        \subfloat[Superset]{\includegraphics[height=3cm]{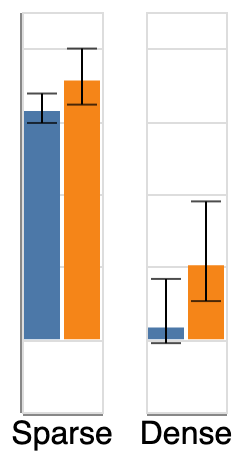}%
            \label{subfig:superset}}%
        \vspace{-.5em}
        \caption{The effect of tile granularity on query time compared to untiled videos. All videos used a one second tile layout duration.
        Objects occupy $<$20\% of each frame on average in ``sparse'', and $\geq$20\% in ``dense'' videos.}
        % In ``sparse'' videos, objects take $<$20\% of each frame on average, while in ``dense'' videos they occupy $\ge$20\%.}
        % \vspace{-1.75em}
        \label{fig:tileGranularityPlot}
    \end{figure}
}

\newcommand{\tileLayoutDurationVsQueryTimePlot}{
    \renewcommand{\FBbskip}{-1em}
    \begin{figure}
        \centering
        \includegraphics[width=0.95\columnwidth]{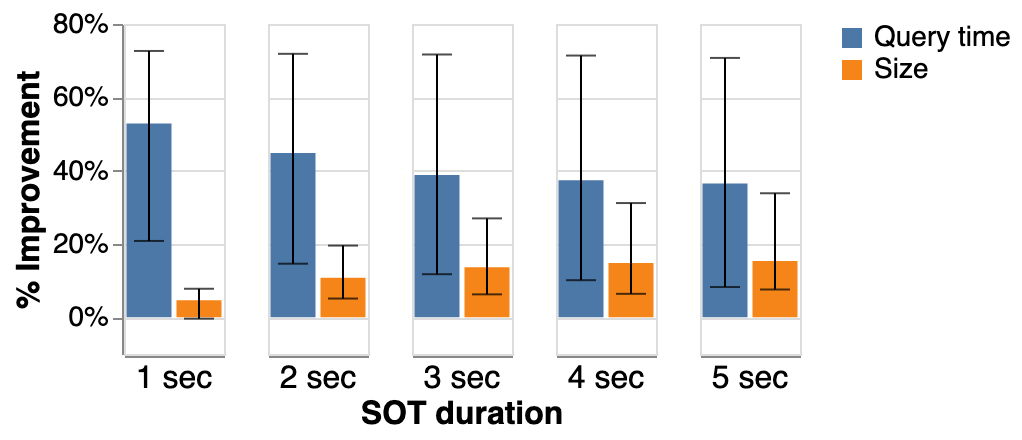}
        \vspace{-1em}
        \caption{This plot shows the effect of SOT duration on query time and storage cost. Tiled videos were encoded with fine-grained tiles and a GOP length equal to the SOT duration.}
        % \vspace{-1.25em}
        \label{fig:tileLayoutDurationPlot}
    \end{figure}
}

\newcommand{\tileImprovementsPlot}{
    \renewcommand{\FBbskip}{-1em}
    \begin{figure}
        \centering
        \subfloat[][]{\includegraphics[width=0.45\columnwidth]{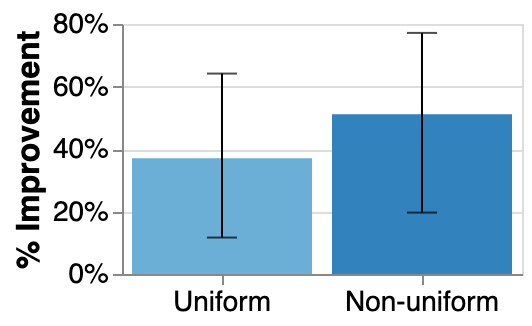}%
            \label{subfig:tilingOnQueryTime}}%
        \hfil
        \subfloat[][]{\includegraphics[width=0.45\columnwidth]{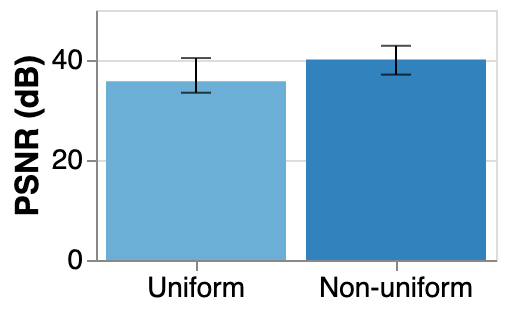}%
            \label{subfig:tilingOnQuality}}%
        % \vspace{-.5em}
        \caption[]{\subref{subfig:tilingOnQueryTime} shows the improvement in query time achieved by tiling the video using the fastest uniform and non-uniform layout for each video and query object. \subref{subfig:tilingOnQuality} shows the quality of these layouts compared to the untiled video.}
        % The dashed line indicates the minimum PSNR value that is considered to have acceptable visual quality.}
        \vspace{-1em}
        \label{fig:tileImprovementsPlot}
    \end{figure}
}

\newcommand{\numUniformTilesPlot}{
    \renewcommand{\FBbskip}{-1em}
    \begin{figure}
        \centering
        \includegraphics[width=0.95\columnwidth]{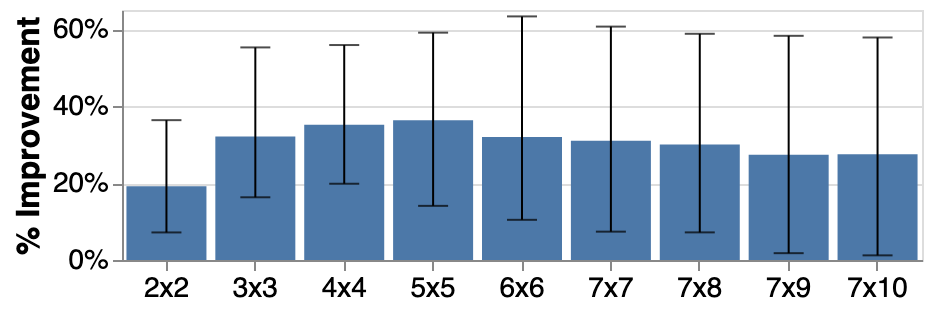}
        \vspace{-1em}
        \caption{This figure shows improvement in query time achieved with various uniform layouts compared to the untiled video.}
        % \vspace{-1.75em}
        \label{fig:numUniformTilesOnQueryTime}
    \end{figure}
}

\newcommand{\tilingHelpfulnessCutoff}{
    \renewcommand{\FBbskip}{-1.75em}
    \begin{figure}
        \centering
        \includegraphics[width=0.95\columnwidth]{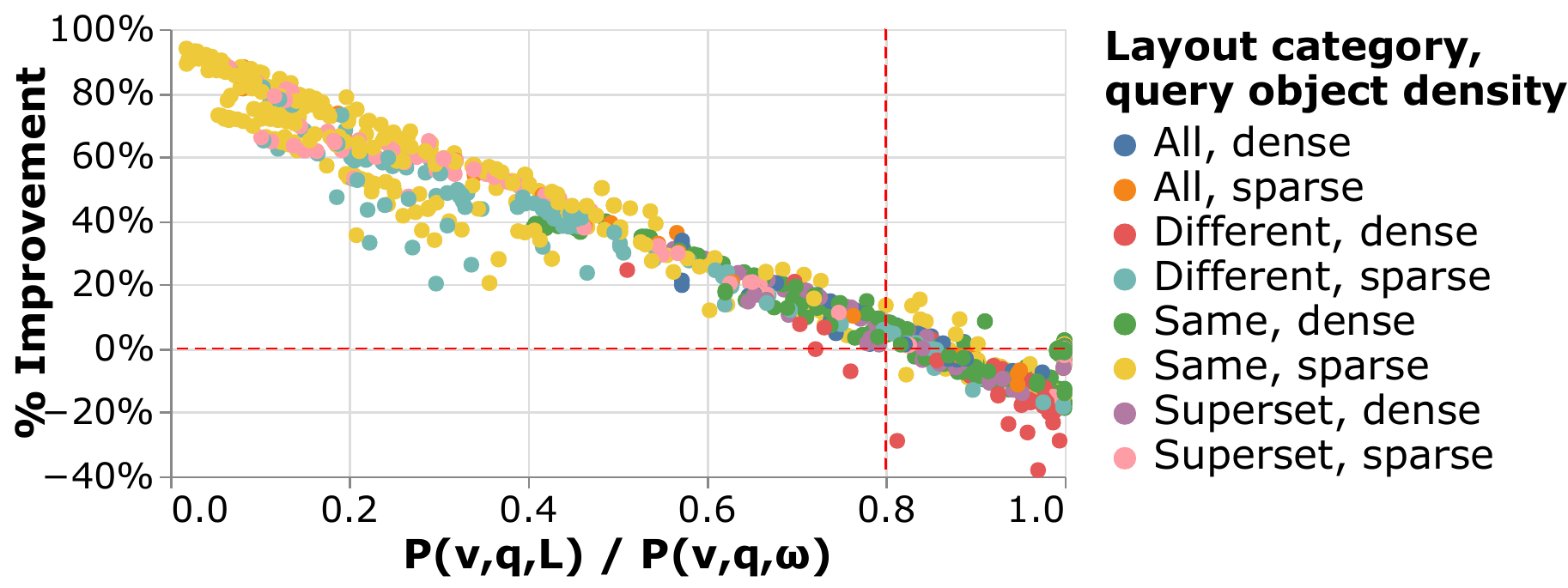}
        \vspace{-.5em}
        \caption{Ratio of the number of pixels decoded with a non-uniform layout to the number decoded without tiles
        % ($P(v, q, L) / P(v, q, \omega)$)
        vs. performance improvement. Each point represents a video, query object, and non-uniform layout. Points below the horizontal line at 0\% represent cases where queries ran more slowly on the tiled video. Points to the right of the vertical line at 0.8 represent videos that would not be tiled when the threshold for tiling requires the ratio to be $ < 0.8$.}
        % \vspace{-1.5em}
        \label{fig:tilingCutoff}
    \end{figure}
}

\newcommand{\cheapTilingPlots}{
    \begin{figure}
        \centering
        \begin{subfigure}{0.45\columnwidth}
            \centering
            \includegraphics[width=\linewidth]{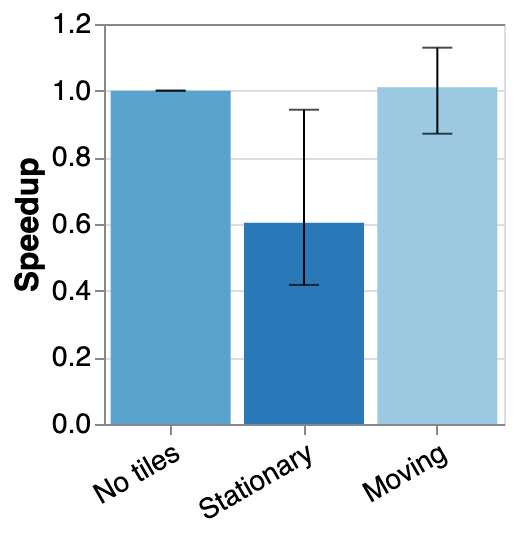}
            \caption{Background subtraction-based tiles}
            \label{subfig:knnTiles}
        \end{subfigure}
        \begin{subfigure}{0.45\columnwidth}
            \centering
            \includegraphics[width=\linewidth]{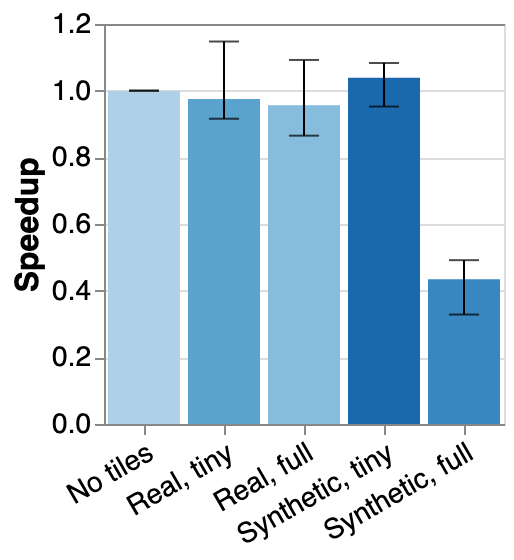}
            \caption{YOLO-based tiles}
            \label{subfig:yoloTiles}
        \end{subfigure}
        \caption{\subref{subfig:knnTiles} shows the effectiveness of tiling around foreground objects detected using KNN-based background subtraction \todo{cite opencv}. It performs better when the camera is stationary. \subref{subfig:yoloTiles} shows the effectiveness of tiling around all objects detected using YOLOv3-tiny run every frame, or full YOLOv3 run every 5 frames. Results are split out based on whether the video is synthetic or not because the synthetic traffic videos contain fewer objects. In many cases, tiling around all of the detected objects actually makes queries slower than not tiling.}
        \label{fig:cheapTiling}
    \end{figure}
}

\newcommand{\workloadFigures}{
    \renewcommand{\FBbskip}{-1.5em}
    \begin{figure}
        \centering
        \subfloat{\includegraphics[width=\columnwidth]{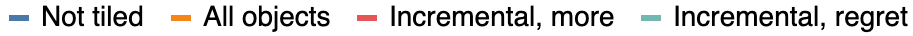}}
        \vspace{-2em}
        \setcounter{subfigure}{0}

        \subfloat[Workload 1]{\includegraphics[width=0.48\columnwidth]{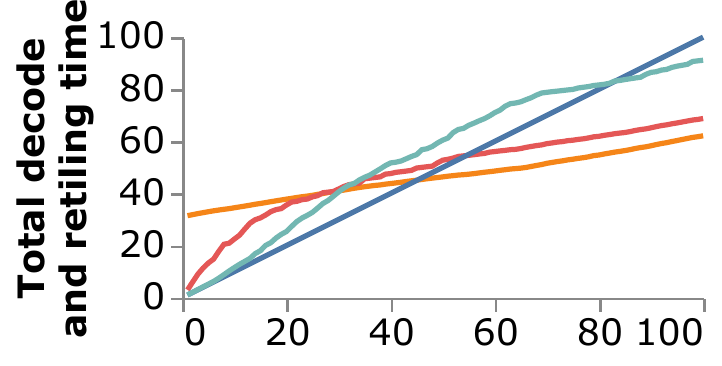}%
            \label{subfig:workload1}}%
        \hfil
        % \begin{subfigure}{0.23\textwidth}
        %     \centering
        %     \includegraphics[width=\linewidth]{figures/unknown_workload/combined_charts/zipf_random_obj}
        %     \caption{Workload 2}
        %     \label{subfig:workload2}
        % \end{subfigure}
        \subfloat[Workload 2]{\includegraphics[width=0.41\columnwidth]{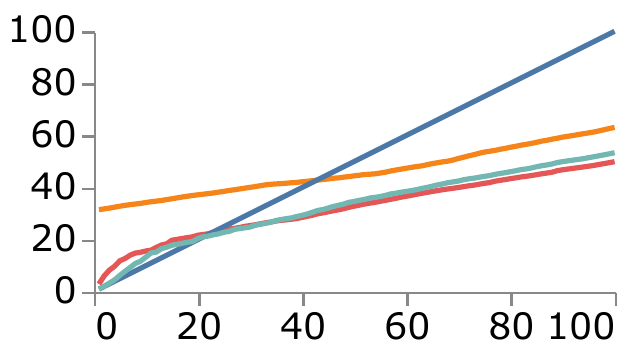}%
            \label{subfig:workload2}}
        \vspace{-1em}

        \subfloat[Workload 3]{\includegraphics[width=0.48\columnwidth]{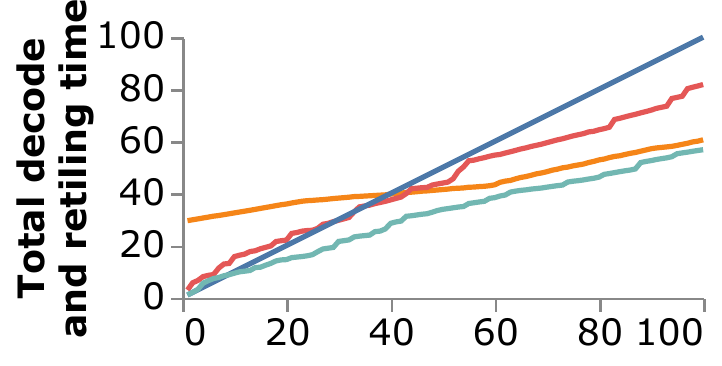}%
            \label{subfig:workload3}}%
        \hfil
        \subfloat[Workload 4]{\includegraphics[width=0.41\columnwidth]{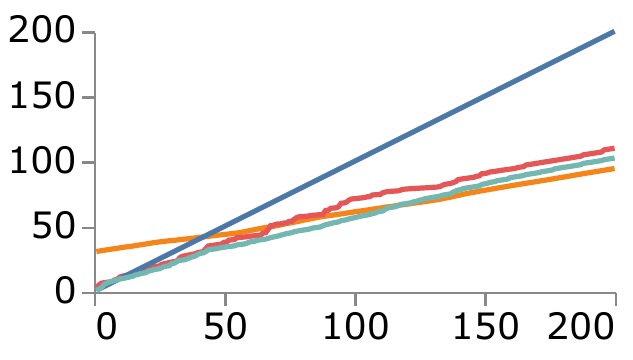}%
            \label{subfig:workload4}}%
        \vspace{-1em}
        % \begin{subfigure}{0.23\textwidth}
        %     \centering
        %     \includegraphics[width=\linewidth]{figures/unknown_workload/combined_charts/window5_objPerSection_obj}
        %     \caption{Workload 6}
        %     \label{subfig:workload6}
        % \end{subfigure}

        \subfloat[Workload 5]{\includegraphics[width=0.48\columnwidth]{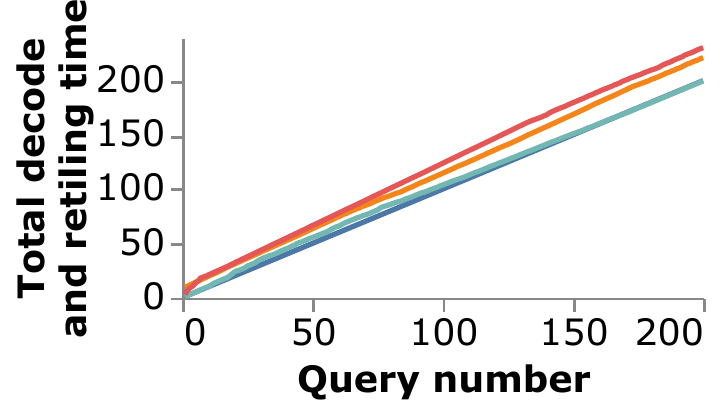}%
            \label{subfig:workload5}}
        \hfil
        \subfloat[Workload 6]{\includegraphics[width=0.41\columnwidth]{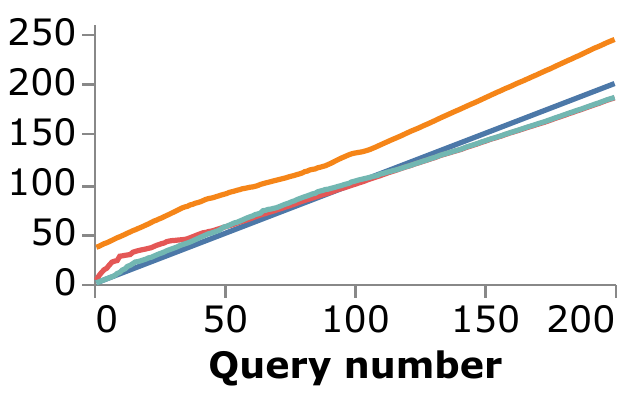}%
            \label{subfig:workload6}}%
        \vspace{-.5em}
        \caption{
            % Cumulative decode and re-tiling time for various strategies and workloads. All values are normalized to the time to execute each query over the untiled videos.
            Cumulative decode and re-tiling time for various workloads. Values are normalized to the time to execute each query over the untiled videos.
            }
        %\vspace{-1em}
        \label{fig:visualroadWorkloads}
    \end{figure}
}

\newcommand{\workloadCostWithDetection}{
    \begin{figure}
        \centering
        \includegraphics[width=0.95\columnwidth]{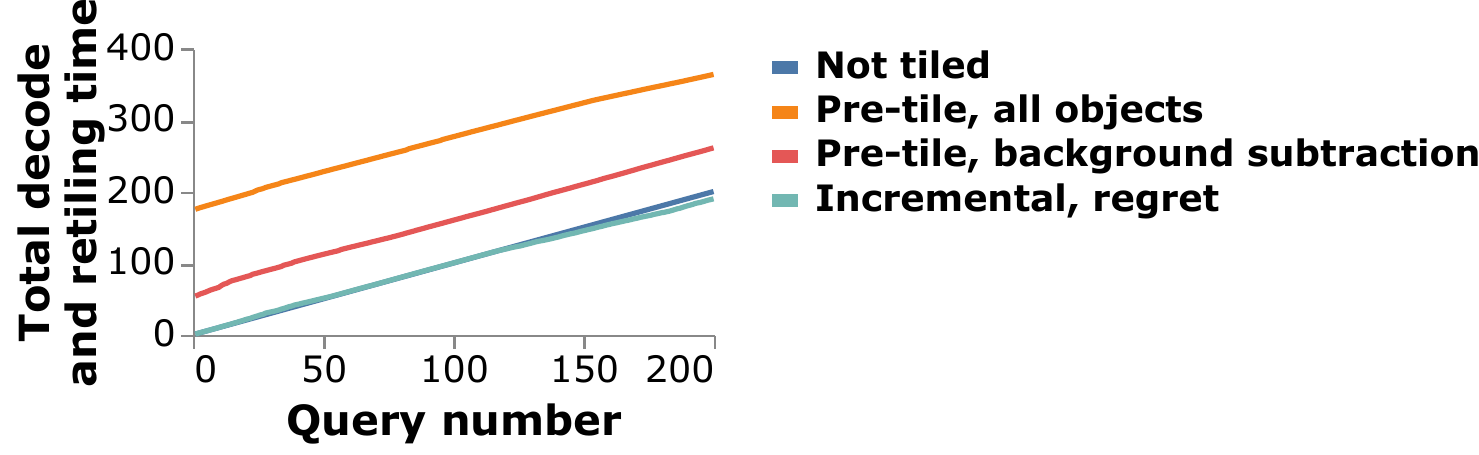}
        % \vspace{-.5em}
        \caption{Cumulative decode and re-tiling time, including initial detection costs. All values are normalized to the time to execute each query over the untiled videos.}
        \vspace{-1em}
        \label{fig:workloadCostWithDetection}
    \end{figure}
}

\newcommand{\tileEveryFivePlot}{
    \begin{figure}
        \centering
        \includegraphics[width=0.7\columnwidth]{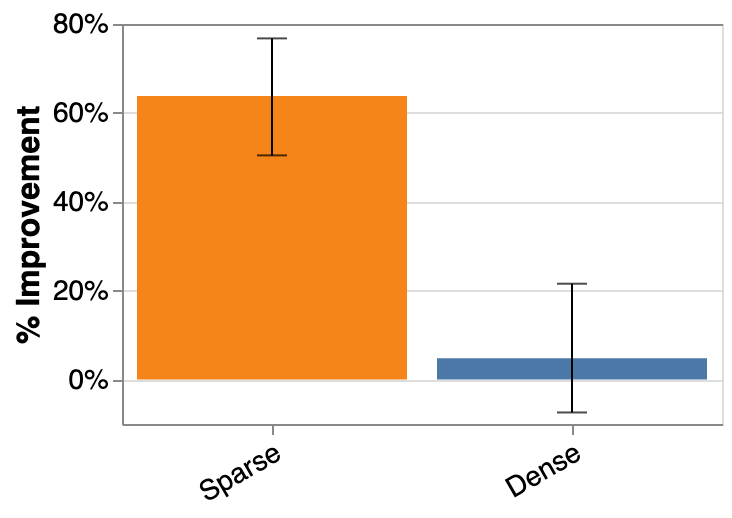}
        \caption{Improvement in query time achieved with fine-grained tile layouts around all objects detected every 5 frames. In ``sparse'' videos, detected objects take up less than 20\% of each frame on average, while in ``dense'' videos they take up at least 20\%. \brandon{Can probably move to text.}}
        \label{fig:tileEveryFive}
    \end{figure}
}

\newcommand{\systemOverview}{
    \renewcommand{\FBbskip}{-1.5em}
    \begin{figure}[t!]
        \centering
        \includegraphics[width=0.95\columnwidth]{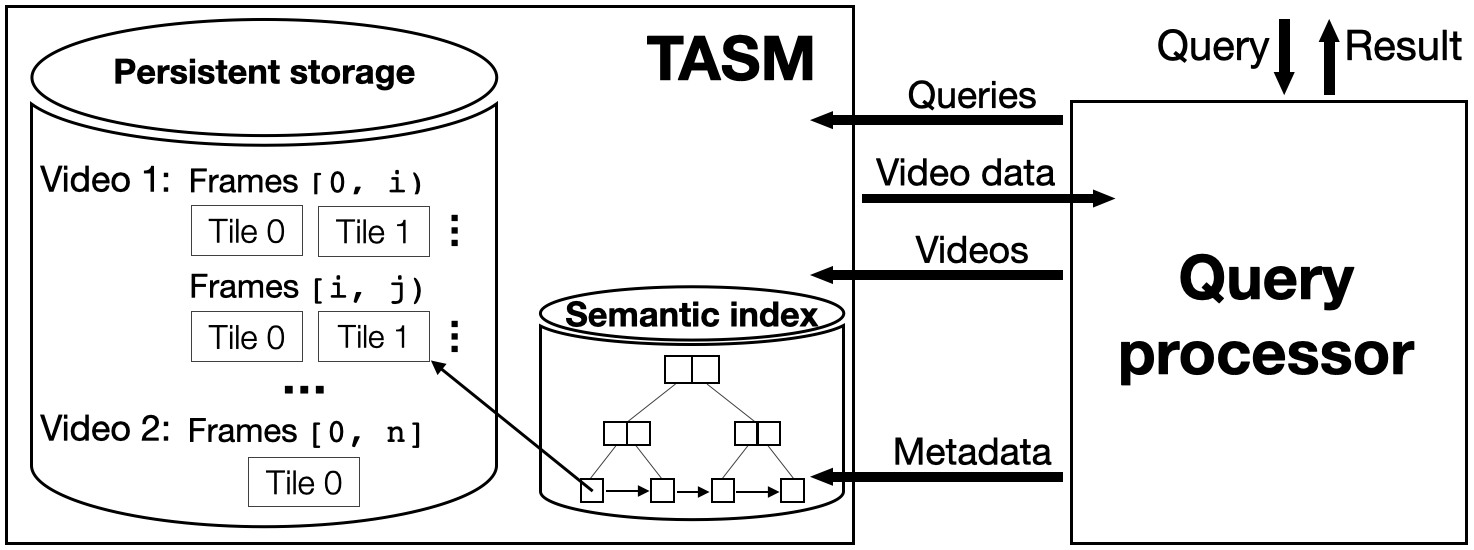}
        \vspace{-.5em}
        \caption{Overview of how TASM integrates with a VDBMS.}
        \label{fig:systemOverview}
    \end{figure}
}

\newcommand{\eagerVsLazy}{
    \begin{figure}
        \centering
        \includegraphics[width=0.95\columnwidth]{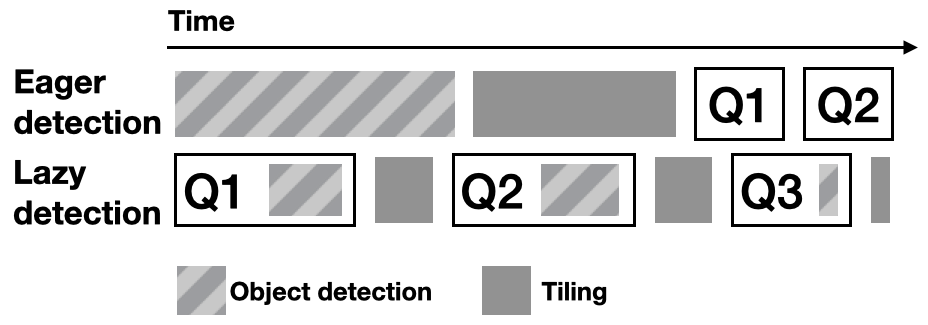}
        \vspace{-1.5em}
        \caption{Example progress of eager and lazy tiling strategies. Queries initially take longer with the lazy strategy, but as objects are detected and sections of the video are encoded with tiles, query times improve.}
        \vspace{-1.5em}
        \label{fig:eagerVsLazy}
    \end{figure}
}

\newcommand{\objDetectionResultsCombined}{
    \renewcommand{\FBbskip}{-1.5em}
    \begin{figure}\TopFloatBoxes
        % \captionsetup{justification=raggedright}
        \begin{floatrow}
            \ffigbox{
                \includegraphics[width=0.95\columnwidth]{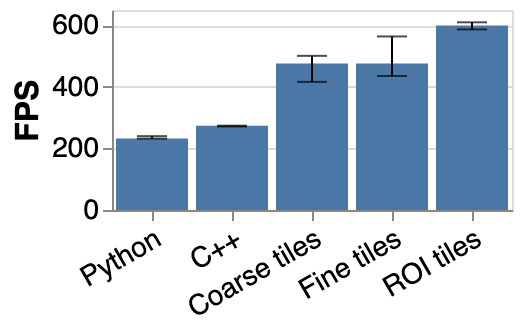}}
                {\vspace{-1em}
                \caption{\revisioncolor{Specialized model preprocessing throughput}}
                \label{fig:preprocessingTput}}
            \killfloatstyle\hspace{-1em}\ttabbox
                {
                    \begingroup
                    \vspace{-0.5em}
                    \setlength{\tabcolsep}{2pt}
                    \begin{tabular}{c c c c}
                        \toprule
                                    & Day 1 & Day 2 & Day 3 \\
                            \hline
                            Full    & 0.79  & 0.51  & 0.56 \\
                            ROI     & 0.84  & 0.61  & 0.51 \\
                            % Coarse-original  & 0.79  & 0.67  & 0.58 \\
                            Coarse  & \revisioncolor{0.76}  & \revisioncolor{0.60}  & \revisioncolor{0.54} \\
                        \bottomrule
                    \end{tabular}
                    \endgroup
                } 
                {\caption{Model accuracy}
                \label{table:Accuracy}}
        \end{floatrow}
    \end{figure}
    % Absolute error results:
    %           & Day 1 & Day 2 & Day 3 \\
    % Full      & 0.22  & 0.46  & 0.42 \\
    % ROI       & 0.16  & 0.38  & 0.45 \\
    % Coarse    & 0.21  & 0.31  & 0.41 \\
    % Coarse-G  & 0.25  & 0.38  & 0.43 \\
}

\newcommand{\visualRoadBenchmarkFigure} {
    \renewcommand{\FBbskip}{-.5em}
    \begin{figure}
        \centering
        \includegraphics[width=0.95\columnwidth]{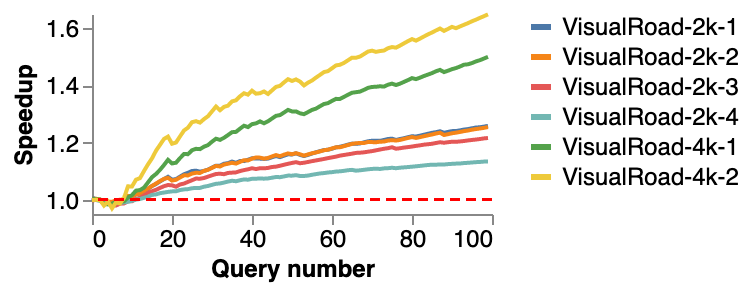}
        \vspace{-1em}
        \caption{\revision{R1W3}{Speedup achieved with TASM over the Visual Road object detection workload. The lines show the median speedup over six orderings of the queries.}}
        \label{fig:visualRoadBenchmarkFigure}
    \end{figure}
}

%!TEX root = paper.tex

\newcommand{\videoSetsTable}{
    \begingroup
    \setlength{\tabcolsep}{2pt}
    \renewcommand{\FBbskip}{-1.5em}
    \begin{table}
        \caption{Video datasets}
        \vspace{-.5em}
        \centering
        \footnotesize
        \begin{tabular}{ l l l l l l }
            \toprule
            Video dataset   & Duration  & Res.      & Per-frame         & Frequently \\
                        &        (sec.)  &          & object             & occurring objects \\
                        &                 &         & coverage (\%)   & \\
            \hline
            Visual Road~\cite{DBLP:conf/sigmod/HaynesMBCC19}\textsuperscript{\textdagger} & 540--900 & 2K, 4K & 0.06--10 & car, person \\[2pt]
            Netflix public~\cite{netflixPublicDataset} & 6 & 2K  & 0.32--49 & person, car, bird \\[2pt]
            Netflix OS~\cite{netflixOpenSourceAssets}\textsuperscript{*} & 720 & 2K, 4K & 25--45 &  person, car, sheep \\[2pt]
            XIPH~\cite{xiphDataset} & 4--20 & 2K, 4K & 2--59 & car, person, boat\\[2pt]
            MOT16~\cite{DBLP:journals/corr/MilanL0RS16} & 15--30 & 2K & 3--36 & car, person \\[2pt]
            El Fuente~\cite{cdvlElFuente} & 480 (full) & 4K & 1--47 & person, car, \\[2pt]
            & \multicolumn{2}{l}{15--45 (scenes)} & & boat, bicycle \\
            \bottomrule
            \textsuperscript{\textdagger} Synthetic videos & \multicolumn{3}{l}{\textsuperscript{*} Both real and synthetic videos}
        \end{tabular}
        % \vspace{-1.5em}
        \label{table:videoSets}
    \end{table}
    \endgroup
}

\newcommand{\workloadResultsTable}{
    \begingroup
    \scriptsize
    \setlength{\tabcolsep}{2pt}
    \renewcommand{\FBbskip}{-1.5em}
    \begin{table}[t!]
        \centering
        % \vspace{-1em}
        \caption{Cumulative workload time. Values are normalized to the time to execute each query over untiled videos.}
        \vspace{-1em}
        \centering
        \begin{tabular}{ c | c | c c c | c c c | c c c }
            \toprule
            % Workload
              & Not & \multicolumn{3}{|c|}{All objects} & \multicolumn{3}{|c|}{Incremental,} & \multicolumn{3}{|c}{Incremental,} \\
            &  tiled & \multicolumn{3}{|c|}{} & \multicolumn{3}{|c|}{more} & \multicolumn{3}{|c}{regret} \\
            &   & 25\% & 50\% & 75\%  & 25\% & 50\% & 75\%    & 25\% & 50\% & 75\% \\
            \hline
            W1  & 100   & 60 & 65 & 75    & 62 & 69 & 71    & 88 & 91 & 96 \\
            W2  & 100   & 59 & 67 & 70    & 40 & 50 & 57    & 42 & 53 & 60 \\
            W3  & 100   & 62 & 64 & 65    & 74 & 82 & 85    & 54 & 57 & 67 \\
            W4  & 200   & 81 & 102 & 111  & 101 & 110 & 121 & 100 & 103 & 124 \\
            W5  & 200   & 205 & 221 & 238 & 192 & 230 & 249 & 187 & 200 & 214 \\
            W6  & 200   & 203 & 244 & 261 & 135 & 186 & 250 & 147 & 186 & 219 \\
            \bottomrule
        \end{tabular}
        % \vspace{-1.5em}
        \label{table:workloadResults}
    \end{table}
    \endgroup
}

\newcommand{\apiTable}{
    \vspace{-1em}
    \begin{table}[h!]
        \centering
        % \caption{\revision{R3W1}{TASM API}}
        % \vspace{-.5em}
        \begin{tabular}{l l l}
            \toprule
            \revisioncolor{Method} & \revisioncolor{Parameters} & \revisioncolor{Result} \\
            \midrule

            $\textsc{Scan}$         & $video,\, L: labels,\, T: times$            & Pixel[]\\
            $\textsc{AddMetadata}$  & $video,\, frame,\, label,$                  & --- \\
                                    & $x_1,\, y_1,\, x_2,\, y_2$                  & \\

            \bottomrule
        \end{tabular}
        \vspace{-1em}
        \label{table:api}
    \end{table}
}

\newcommand{\incrementalLayoutPseudocode} {
    \renewcommand{\FBbskip}{-2em}
    \begin{figure}[t]
        \small
        \begin{algorithmic}[1]
            \State $O_{Q'} \gets \emptyset$, $L_{alt} \gets \emptyset$, $\forall s_j \in v: \delta^j \gets 0, L_0^j \gets \omega$
            \ForAll {$q_i \in Q$}
                \State $O_{Q'} \gets O_{Q'} \cup O_{q_i}$
                \State $L_{alt}' = \mathcal{P}(O_{Q'})$
                \ForAll {$L_k \in L_{alt}' - L_{alt}$}
                    \For{$m = 0,\ldots,i-1$}
                        \State $\forall s_j \in v: \delta_k^j \gets \delta_k^j + \Delta(q_m, L_m^j, L_k)$
                    \EndFor
                \EndFor
                \State $L_{alt} \gets L_{alt}'$
                \ForAll {$L_k \in L_{alt}$}
                    \State $\forall s_j \in v: \delta_k^j \gets \delta_k^j + \Delta(q_i, L_i^j, L_k)$
                \EndFor
                \ForAll {$s_j \in v$}
                    \State $k^* \gets arg\,max_k \delta_k^j$
                    \If {$\delta_{k^*}^j > \eta \cdot R(s_j, L_{k^*})$}
                        \State Retile $s_j$ with $L_{k^*}$. $\delta^j \gets 0$
                    \EndIf
                \EndFor
            \EndFor
        \end{algorithmic}
        \vspace{-.5em}
        \caption{Pseudocode for incrementally adjusting layouts}
        \label{alg:incrementalLayout}
        % \vspace{-2em}
    \end{figure}
}

%!TEX root = paper.tex

\section{Introduction} \label{sec:introduction}

%Large amounts of video are being captured thanks to the proliferation of cameras,

The proliferation of inexpensive high-quality cameras coupled with recent advances in machine learning and computer vision have enabled new applications on video data such as automatic traffic analysis~\cite{DBLP:conf/cloud/LuCK16,DBLP:conf/nsdi/ZhangABPBF17}, retail store planning~\cite{DBLP:journals/pvldb/KangBZ19}, and drone analytics~\cite{DBLP:conf/mobicom/WangCC16, DBLP:conf/edge/WangFCGBPYS18}. This has led to a class of database systems specializing in video data management that facilitate query processing over videos~\cite{Panorama:pvldb/ZhangK19,DBLP:journals/pvldb/KangEABZ17,Scanner:tog/PomsCHF18,Focus:osdi/HsiehABVBPGM18,VStore:eurosys/XuBL19,DBLP:journals/pvldb/KangBZ19}.

% <original>
% The amount of video data being created is rapidly growing\brandon{Ideally don't repeat the first sentence of the abstract}. YouTube alone has more than 400 hours of video uploaded every minute~\cite{youtube-stats}\brandon{Since TASM isn't really useful for YouTube video, can we sub something more relevant here?}. At the same time, advances in machine learning and computer vision have enabled new applications on video data such as automatic traffic analysis~\cite{DBLP:conf/cloud/LuCK16,DBLP:conf/nsdi/ZhangABPBF17}, retail store planning~\cite{BlazeIt:corr/abs-1805-01046}, and drone analytics~\cite{DBLP:conf/mobicom/WangCC16, DBLP:conf/edge/WangFCGBPYS18}. This has led to a class of database systems specializing in video data management that facilitate query processing over videos~\cite{Scanner:tog/PomsCHF18,Focus:osdi/HsiehABVBPGM18,VStore:eurosys/XuBL19,Panorama:pvldb/ZhangK19,DBLP:journals/pvldb/KangBZ19,DBLP:journals/pvldb/KangEABZ17}.
% </original>

%For some queries this requires a \textit{full scan} through the entire video.
%Efficient storage managers for video data are lacking, despite this preprocessing step being expensive.

A query over a video comprises two steps. First, read the video file from disk and decode it. Second, process frames to identify and return pixels of interest or compute an aggregate.
Most systems, so far, have focused on accelerating and optimizing the second step~\cite{Focus:osdi/HsiehABVBPGM18, VStore:eurosys/XuBL19, DBLP:journals/pvldb/KangBZ19, DBLP:conf/sigmod/BastaniHBGABCKM20}, often assuming that the video is already
decoded and stored in memory~\cite{DBLP:journals/pvldb/KangBZ19,DBLP:journals/pvldb/KangEABZ17,DBLP:conf/sigmod/LuCKC18}, which is not feasible in practice.
% \dong{This paragraph can be more clear. The two steps are a bit unclear, and readers might be confused about which operations belong to which step. It’d be worth making clear that we’re proposing a storage manager that optimizes the “reading and decoding” before mentioning that the two types of queries that can benefit from this, otherwise the reader may wonder what exactly does the storage manager do.}

The lack of efficient storage managers in existing video data management systems significantly impacts queries. First, \textit{\revision{R3D3}{subframe selection}} queries (e.g., 
% ``Show me the previously identified video snippets of hummingbirds feeding on honeysuckles''
``Show me video snippets \revision{R3D3}{cropped to show} previously identified hummingbirds feeding on honeysuckles''
% \enhao{The definition of subframe selection is unclear to me. Does subframe selection select full frames or regions of frames that only contain hummingbirds?}
% \brandon{is the `previously identified' part important here?} 
% \brandon{As phrased, this sounds more like content retrieval than subframe selection}
% \maureen{Good point. I tried to make it sound more like subframe selection, but this may need some iteration.}
) are common and their execution bottleneck is at the storage layer since these queries are selections, reading and returning pixels without additional operations. Second, \textit{object detection} queries, which extract new semantic information from a video (e.g., ``Find all sightings of hummingbirds in this new video'') require the execution of expensive deep learning models. To avoid applying such models to as many frames as possible, query plans typically include an initial \textit{full scan} phase that applies a cheap predicate~\cite{DBLP:conf/sigmod/LuCKC18} or a specialized model~\cite{DBLP:journals/pvldb/KangEABZ17} to the entire video to filter uninteresting frames. The overhead of reading and decoding the video file is known to significantly hurt the performance of this phase~\cite{DBLP:journals/corr/abs-2007-13005}.

In this paper, we introduce TASM, a storage manager that greatly improves the performance of \revision{R3D3}{subframe selection} queries and the full scan phase of object detection queries by providing spatial random access within videos.
TASM exploits the observation that objects in videos frequently lie in subregions of video frames.
For example, a traffic camera may be oriented such that it partially captures the sky, so vehicles only appear in the lower portion of a frame.
% An amber alert application only needs to analyze the part of the frame containing vehicles. 
% Similarly, birds tend to appear near a bird-feeder, so an ornithologist looking for hummingbirds could focus their search on the area around the feeder. 
\revision{R1W2}{
Analysis applications such as running license plate recognition~\cite{DBLP:conf/cloud/LuCK16} or extracting image patches for vehicle type recognition~\cite{DBLP:conf/cloud/LuCK16} only need to operate on the parts of the frame containing vehicles.
Privacy applications such as blurring license plates and faces~\cite{waymoDatasetFAQ} or performing region of interest-based encryption~\cite{DBLP:conf/eusipco/AbuTahaSHDVV18} similarly only need to modify the parts of the frame that contain sensitive objects.
% Subframe selection enables the system to decode and process only the interesting parts of the frame, speeding up these types of operations.
} 
% \magda{Good examples and nice citations, be should combine the new examples with the older examples above. Otherwise it's too many examples. Then we should clarify how subframe selection helps in those examples. Something like: ``In analysis tasks, subframe selection enables the system to only decode and modify the tiles of interest, speeding up these types of operations.''}

\tileAndFilesFigure

Using its spatial random access capability, TASM enables reading from disk and decoding only the parts of the frame that are interesting to queries. Providing such a capability is difficult because the video encoding process introduces spatial and temporal dependencies within and between frames.  To address this problem, TASM subdivides video frames into smaller pieces called \textit{tiles} that can be processed independently.
% \magda{Also refer to Figure 1. Move Figure 1 closer to here. And perhaps show an actual frame in Figure 1 split into tiles?}
\revision{R3W2}{As shown in \cref{fig:tileAndFiles}, each tile contains a rectangular subregion of the frame that can be decoded independently because there are no spatial dependencies between tiles. In contrast, current state of the art incurs the cost of decoding entire frames.}
% \brandon{Do we want to draw a contrast here with current state of the art, where the cost of decoding the entire frame must be incurred?  Readers might not yet be primed to appreciate what a `spatial dependency' is.}}
TASM optimizes how a video is divided into tiles and stored on disk to reduce the amount of work spent decoding and preprocessing 
parts of the video not involved in a query.
Through its use of tiles, TASM implements a new type of optimization that we call \textit{semantic predicate pushdown} where predicates are pushed below the decoding step and only tiles of interest are read from disk, decoded, and processed.

Building TASM raises three challenges. The first challenge is fundamental, but important: Given a video file with known semantic content (i.e., known object locations within video frames) and a known query workload, TASM must decide on the optimal tile layout, choosing from among layouts with uniform or non-uniform tiles and either fine-grained or coarse-grained tiles. TASM must also decide whether different tile layouts should be used in different parts of a video.
To do this effectively, TASM must accurately estimate the cost of executing a query with a given tile layout.
TASM therefore drives its selection using a cost function that balances the benefits of processing fewer pixels against the overhead of processing more tiles for a given tile layout, video content, and query workload.
In this paper, we experimentally demonstrate that non-uniform, fine-grained tiles outperform the other options. Additionally, we find that optimizing the layout for short sections of the video (i.e., every 1~second) maximizes query performance with no storage overhead. Given a video file, TASM thus splits it into 1~second fragments and selects the optimal fine-grained tile layout for each fragment. 

%Additionally, TASM uses different layouts for different regions of a video in order to minimize the cost of a workload.

%It estimates the retiling cost and the benefit of different tile layouts for a given query workload and uses these cost estimates to select the optimal tiling for a given query workload.

%given knowledge of the semantic content of a video and objects targeted by content retrieval queries, TASM must determine if tiling will be beneficial to query execution and, if tiling is likely to be beneficial, then optimize the layout of tiles to accelerate queries.  TASM addresses these challenges by introducing a cost function to determine the cost-effectiveness of tiling a video fragment. 

% \magda{Here, we need to say what is interesting/hard/non-obvious about the cost function}. 

% \brandon{(3) maintaining detection quality?}\brandon{I wonder if a new reader will have enough context to understand this sentence.  Should we say a little more about `how a video is tiled' before going into this?  Maybe tie back to an example?  Not sure.}.
% \brandon{Is it clear to the reader that there are 99999999999 possible tilings?}.

%The second challenge is that TASM will not know where objects are located nor which objects content retrieval queries will target when the video is initially stored.
%As a result, TASM must incrementally tile and retile a video in a way that yields high query performance with low overhead as it learns about the video contents and workload.

The second challenge is that the semantic content and the query workload for a video are typically discovered over time as users execute object detection and \revision{R3D3}{subframe selection} queries.
TASM therefore lacks the information it needs to design optimal tile layouts.
To address this challenge, TASM incrementally updates a video's tile layout as queries to detect and retrieve objects are executed.  TASM uses different tile layouts in different parts of the video, and independently evolves the tile layout in each section. To do this, TASM builds on techniques from database cracking~\cite{DBLP:conf/cidr/IdreosKM07, DBLP:journals/pvldb/HalimIKY12} and online indexing~\cite{DBLP:conf/icde/BrunoC07}. To decide when to re-tile portions of the video and which layout to use, TASM maintains a limited set of alternative layouts based on past queries. It then uses its cost function to accumulate estimated performance improvements offered by these tile layouts as it observes queries. Once the estimated improvement, also called regret~\cite{DBLP:conf/icde/DashKA09}, of a new layout offsets the cost of reorganization, TASM re-tiles that portion of the video. By observing multiple queries before making tiling decisions, TASM designs layouts optimized for multiple query types.  For the ornithology example, TASM could tile around hummingbirds \textit{and} flowers that are likely to attract them.

% but it still improves query performance by adapting the layout as it observes queries and incrementally learns about the semantic content of the video.

% Incremental tiling, however, can be expensive and even slow down query execution. For example, if TASM tiles part of a video around cars but the next query retrieves pedestrians, the tile layout designed around cars could cause the query for pedestrians to execute more slowly than if the video were not tiled at all. Reorganizing a video with a new tile layout is expensive, so picking a bad layout can not only slow down future queries, but also waste time spent on encoding.
% For example, TASM could tile around cars \textit{and} pedestrians to speed up queries for both objects\brandon{Can probably do some thinning in this paragraph}.

% While the number of content retrieval queries over a video may be much larger than the number of object detection queries, object detection is still extremely expensive and therefore important to optimize.

The third challenge lies in the initial phase that identifies objects of interest in a new video. This phase is both expensive and requires at least one full scan over the video, generally using a cheap model to filter frames or compute statistics.
The models used in the full scan phase are limited by video decoding and preprocessing throughput~\cite{DBLP:journals/corr/abs-2007-13005}.
To address this final challenge, TASM uses semantic predicate pushdown where the semantic predicate is not a specific object type, but rather a general region of interest (ROI).
TASM bootstraps an initial tile layout using an inexpensive predicate that identifies ROIs within frames.
This predicate can use background segmentation to find foreground objects, motion vectors to identify areas with large amounts of motion, or even a specialized neural network designed to identify specific object types. When an object detection query is executed, TASM only decodes the tiles that contain ROIs, hence filtering regions of the frame before the decode step.
TASM thus alleviates the bottleneck for the full scan phase of object detection queries by reducing the amount of data that must be decoded and preprocessed.
TASM can be directly incorporated into existing techniques and systems that accelerate the extraction of semantic information from videos (e.g., ~\cite{DBLP:journals/pvldb/KangBZ19, DBLP:conf/sigmod/BastaniHBGABCKM20}).

In summary, the contributions of this paper are as follows:
\begin{itemize}
    \setlength\topsep{0pt}
    \setlength\itemsep{1pt}
	\setlength{\parskip}{0pt}
    \setlength{\parsep}{0pt}

    \item We develop TASM\footnote{Code is available at \url{https://github.com/uwdb/TASM}.}, a new type of storage manager for video data that splits video frames into independently queryable tiles. TASM optimizes the tile layout of a video file based on its contents and the query workload. By doing so, TASM accelerates queries that retrieve objects in videos while keeping storage overheads low and maintaining good video quality.
    \item We develop new algorithms for TASM to dynamically evolve the video layout as information about the video content and query workload becomes available over time.
    \item We extend TASM to cheaply profile videos and design an initial layout around ROIs when a video is initially ingested. This initial tiling reduces the preprocessing work required for object detection queries.
    \end{itemize}
    
We evaluate TASM on a variety of videos and workloads and find that the layouts picked by TASM speed up \revision{R3D3}{subframe selection} queries by an average of 51\% and up to 94\% while maintaining good quality, and that TASM automatically tunes layouts after just a small number of queries to improve performance even when the workload is unknown. We also find that TASM improves the throughput of the full scan phase of object detection by up to $2\times$ while maintaining high accuracy.

\section{Background}
\label{sec:videoencoding}

%\brandon{Probably won't have space to cover all of these things.  May need to define some terms, spend some time on tiling, and have pointers to other resources.  Probably a figure for tiling.}

%\subsection{Video data management systems}

%In this section, we motivate the functionality of our proposed storage manager and provide important background on video encoding.

%\subsection{Motivation}

%\magda{This subsection should include a motivating application with example queries that will justify our tile-based storage manager later. Ideally, we should also have a toy example application that we will use throughout Sections 3 and 4.}

%\subsection{Video encoding} \label{sec:videoencoding}

%\brandon{Could cite specs if you want to}

Videos are stored as encoded files due to their large size. Video codecs such as H264~\cite{H264Specification}, HEVC~\cite{HEVCSpecification}, and AV1~\cite{AV1Specification} specify 
% encoding and decoding
algorithms used to (de)compress videos. While the specific algorithms used by various codecs differ, the high-level approach is the same
as we describe in this section.

%Frames are grouped temporally into groups of pictures that provide temporal random access points, and a feature of modern video codecs called ``tiles'' can be used to introduce spatial random access points.

\textbf{Groups of pictures:} A video consists of a sequence of frames, where each frame is a 2D array of pixels. Frames in the sequence are partitioned into \textit{groups of pictures} (GOPs). Each GOP is encoded independently from the other GOPs and is typically one second in duration. The first frame in a GOP is called a \textit{keyframe}. Keyframes allow GOPs to act as temporal random access points into the video because it is possible to start decoding a video at any keyframe. To retrieve a specific frame, the decoder begins decoding at the closest keyframe preceding the frame being retrieved. Keyframes have large storage sizes because they use a less efficient form of compression than other types of frames, so the number of keyframes impacts a video's overall storage size. Videos stored with long GOPs are smaller in size than videos stored with short GOPs, but they also have fewer random access opportunities.
% does not have to start at the beginning of the video; it 

% \brandon{Need a figure if space.}
% \label{sec:backgroundTile}

% \tileAndFilesFigure

\textbf{Tiles:} Compressed videos do not generally support decoding spatial regions of a frame. The encoding process creates spatial dependencies within a frame, and decoders must resolve these dependencies by decoding the entire frame, even if just a small region is requested.  Modern codecs, however, provide a feature called \textit{tiles} that enables splitting frames into independently-decodable regions. \cref{fig:tileAndFiles} illustrates this concept.
% \magda{I reworded the explanation. Not sure if it's clearer, though.}
\revision{R3D4}{
    % Each frame is thus a 2D array of pixels and the video is a sequence of frames. Each frame can be divided into smaller pieces, called tiles, that can be accessed independently. A tile is thus a portion of a frame.
Like frames, tiles are also 2D arrays of pixels. However, a tile only contains the pixels for a rectangular portion of the frame. The full frame is recovered by combining the tiles.
}
Tiles introduce spatial random access points for decoding. To decode a region within a frame, only the tiles that contain the requested region are processed. This flexibility to decode spatial subsets of frames comes with tradeoffs in quality; tiling can lead to artifacts appearing at the tile boundaries~\cite{DBLP:journals/tcsv/SullivanOHW12}, which reduces the visual quality of videos. As such, carefully selecting tile layouts is important for high-quality query results.
\squeeze{
While tiles act as spatial random access points, temporal random access is still provided by keyframes. Tiles are applied to all frames within a GOP, so decoding a tile in a non-keyframe requires decoding that tile in all frames starting from the preceding keyframe.
}

A \textit{tile layout} defines how a sequence of frames is divided into tiles. A layout $L {=} \left(n_{r}, n_{c}, \{h_1, \ldots, h_{n_r}\}, \{w_1, \ldots , w_{n_c}\}\right)$ is defined by the number of rows and columns, $n_r$ and $n_c$, the height of each row, and the width of each column. These parameters define the $(x, y)$ offset, width, and height of the $n_r {\cdot} n_c$ tiles. An untiled video is a special case of a tile layout consisting of a single tile that encompasses the entire frame: $\omega = \left(1, 1, \{frame\_height\}, \{frame\_width\}\right)$. Valid layouts require tiles to be partitioned along a regular grid, meaning rows and columns extend through the entire frame. We do not consider irregular layouts, which are not supported by the HEVC specification~\cite{HEVCSpecification}. Different tile layouts can be used throughout the video; a \textit{sequence of tiles} (SOT) refers to a sequence of frames with the same tile layout. Changes in the tile layout must happen at GOP boundaries, so every new layout must start at a keyframe. %, which have poor compression performance.
Therefore, changing the tile layout has a high storage overhead for the same reason that starting a new GOP has a high storage overhead.
% This leads to a high storage overhead of changing the tile layout. 
The cost of executing a query over a video encoded with tiles is proportional to the number of pixels and tiles that are decoded.

% The impact on encoding performance is less straightforward; tiles can have both positive and negative effects. Tiles may hurt encoding performance because there are fewer opportunities for compression; the encoder can only search within the same tile rather than the entire frame. However, this limited search space can sometimes actually improve encoding performance because the encoder can more efficiently search within the tile for highly effective compression opportunities [14].

\textbf{Stitching:} Tiles can be stored separately, but they must be combined to recover the original video. Tiles can be combined without an intermediate decode step using a process called \textit{homomorphic stitching}~\cite{LightDB:pvldb/HaynesMABCC18}. Homomorphic stitching interleaves the encoded data from each tile and adds header information so the decoder knows how the tiles are arranged.
% \enhao{Not fully understand the purpose of using stitching. Does stitching combine two encoded tiles into one? But during the query, you still need to decode these tiles eventually, right? Then, what's the difference between ``stitching two encoded tiles, then decoding them as a whole'', and ``decoding two tiles separately, then combining them''?}
%!TEX root = paper.tex

\section{Tile-based storage manager design}

In this section, we present the design of TASM, our tile-based storage manager. TASM is designed to be the lowest layer in a VDBMS. Unlike existing storage managers that serve requests for sequences of frames, TASM efficiently retrieves regions \textit{within} frames to answer queries for specific objects. 
% \sout{By doing so, TASM accelerates queries that request video fragments.}
% TASM is designed for workloads with infrequent but expensive queries to extract high-level semantic information from the video, and multiple content-based retrieval queries for video fragments based on the semantic information\brandon{We should have already covered this in the intro}.

\Cref{fig:systemOverview} shows an overview of how TASM integrates with the rest of a VDBMS. TASM incrementally populates a \textit{semantic index} to store the bounding boxes associated with object detections.
%  and map each bounding box to the tiles that contain it.
% \sout{TASM incrementally populates this semantic index as object detection queries are executed.}
While queries for statistics about the semantic content can use the semantic index to avoid re-running expensive analysis over the frame contents,
% TASM performs two main tasks to accelerate queries for specific objects. First, it incrementally populates the semantic index using object detections that are produced as a byproduct of query execution. Each detection is a bounding box and one or more labels provided by the query processor as metadata. 
TASM uses this index to generate tile layouts, \squeeze{split videos into tiles, store such physically tuned videos as files, and answer content-based queries more efficiently by retrieving only relevant tiles from disk.}

%We first present the API that TASM exposes for the query processor to request video fragments relevant to queries and add metadata to the semantic index in \cref{sec:api}. Sections~\ref{sec:index}-\ref{sec:buildingIndex} then describe TASM's first task: building the semantic index. \Cref{sec:storage} details TASM's second task: designing tile layouts and efficiently retrieving video data to answer queries.

% \brandon{In the previous paragraph I wonder if you want to make the two main tasks performed by the TASM super concrete, e.g. `The TASM performs two major tasks to accelerate answering queries for specific objects.  First, it incrementally populates a semantic index using detected objects that are produced as a byproduct of query execution.  Second, uses this index to answer queries more efficiently.  Then in this paragraph you can say we detail the first task in 3.1-3.4, and section 3.5 the second.'}
% \Cref{sec:api} presents the API TASM exposes for  \Cref{sec:index} describes the semantic index which holds the metadata about the locations of objects within the video, and  \Cref{sec:storage} describes how TASM tiles videos using the semantic index and stores the tiles on disk\brandon{Are the tiles \textit{part} of the semantic index or is this unclustered with leaves that point to tiles?}. Finally, \Cref{sec:query} presents how TASM answers queries over video data.

\systemOverview

\subsection{TASM API}
\label{sec:api}

% \magda{To save space, you can remove the label ``TABLE I'' with the name of the table and include the table as part of the text saying: ``TASM exposes the following access method API: <table>'}

TASM exposes the following access method API: \apiTable

The core method $\textsc{Scan}\left( video, L, T \right)$ \revision{R3W1, D3}{performs subframe selection} by retrieving the pixels that satisfy a CNF predicate on the labels, $L$, and an optional predicate on the time dimension, $T$.
%  while $L = (label=`car \textnormal{'}) \land (label=`red \textnormal{'})$ retrieves pixels belonging to red cars\brandon{Can drop the second example}. 
For example, $L {=} (label{=}`car \textnormal{'}) {\vee} (label{=}`bicycle \textnormal{'})$ retrieves pixels for both cars and bicycles.
% For each disjunctive clause $c {=} l_0 \vee  \cdots \vee l_n$, TASM retrieves pixels that lie in the bounding boxes associated with any $l_i$. For each conjunction, $L {=} c_0 \land \cdots \land c_n$, TASM retrieves pixels that lie in the intersection of bounding boxes associated with all $c_i$. 
% If the predicate $T {=} t_{start} \leq t < t_{end}$ or $T {=} t$ is specified, TASM only considers frames that lie in the specified temporal range.

TASM also exposes an API to incorporate metadata generated during query processing into the semantic index (discussed in the following section). The method $\textsc{AddMetadata}\left( video, frame, label, x_1, y_1, x_2, y_2 \right)$ adds the bounding box $(x_1, y_1, x_2, y_2)$ on $frame$ to the semantic index and associates it with the specified label.

\subsection{Semantic index}
\label{sec:index}

TASM maintains metadata about the contents of videos in a \textit{semantic index}.
% \brandon{This is redundant with the first paragraph of III; can we deduplicate?  Probably a brief mention in the preamble and consolidate the details here}.
The semantic information consists of labels associated with bounding boxes.
Labels denote object types and properties such as color. Bounding boxes locate an object within a frame.
% The search key of the index is a video identifier, a time within the video, a label of interest, and an associated bounding box. 
% The value in the index is a pointer to the underlying tile on disk.
When the query processor invokes TASM's $\textsc{Scan}$ method, TASM must efficiently retrieve bounding box information associated with the specified parameters.
%  labels in $L$ and within the time range specified by $T$.
The semantic index is therefore implemented as a B-tree clustered on $(video, label, time)$. %$\brandon{Maaaaybe drop this if we have to}.
The leaves contain information about the bounding boxes and pointers to the encoded video tile(s) each box intersects based on the associated tile layout.

The semantic index is populated through the \textsc{AddMetadata} method as object detection queries execute. As we discuss in \cref{sec:tilingStrategies}, TASM creates an initial layout around high-level regions of interest within frames to speed up object detection queries. As those queries execute and add more objects to the semantic index, TASM incrementally updates the tile layout to maximize the performance of the observed query workload.

\subsection{Tile-based data storage}
\label{sec:storage}

Having captured the metadata about objects and other interesting areas in a video using the semantic index, the next step is to leverage it to guide how the video data is encoded with tiles.
Two tiling approaches are possible: uniform-sized tiles, or non-uniform tiles whose dimensions are set based on the locations of objects in the video. Both techniques can improve query performance, but tile layouts that are designed around the objects in frames can reduce the number of non-object pixels that have to be decoded. \Cref{fig:comparingCustomLayouts} shows these different tiling strategies on an example frame.
%  that contains cars and pedestrians \dong{The clause can be cut.}.

\comparingCustomLayoutsFigure

\subsubsection{Uniform layouts}

% \magda{This paragraph could be more succinct. It feels like there are repetitions within the paragraph that could be removed}
The uniform layout approach divides frames into tiles with equal dimensions. This approach does not leverage the semantic index, but
% , so the video can be tiled before any metadata is generated.
if objects in the video are small relative to the total frame size, they will likely lie in a subset of the tiles.
% Therefore, queries to retrieve objects from the video can be executed by decoding just a subset of tiles. 
However,
% because the tile layout does not consider the locations of objects in the video\brandon{repetitive},
an object can intersect multiple tiles,
% with part of the object in each tile
as shown in \cref{subfig:uniformLayoutCarPed} where part of the person lies in two tiles. While TASM decodes fewer pixels than the entire frame, it still must process many pixels that are not requested by the query. 
% Moreover, because the tile layouts do not consider the locations of objects, tile boundaries could intersect objects and degrade their visual quality.
Further, the visual quality of the video is reduced because in general a large number of uniform tiles are required to improve query performance, as shown in \cref{subfig:tilingOnQuality}.
% \dong{Since our focus is not on uniform layouts, this subsubsection can be cut a bit.}

% \queryToTilesToDecodeFigure

\fineVsCoarseGrainFigure

\subsubsection{Non-uniform layouts} \label{sec:nonUniformLayouts}

TASM creates non-uniform layouts with tile dimensions such that objects targeted by queries lie within a single tile.
There are a number of ways a given tile layout can benefit multiple types of queries.
If a large portion of the frame does not contain objects of interest, the layout can be designed such that this region does not have to be processed.
If objects of interest appear near each other, a single tile around this region benefits queries for any of these objects. If objects are not nearby but do appear in clusters, creating a tile around each cluster can also accelerate queries for these objects.

\Cref{fig:fineVsCoarseTiles} shows examples of non-uniform layouts around cars. 
%For a given video , set of objects $O$, a sequence of tiles (SOT; see \cref{sec:videoencoding}) from frames $[f_a, f_b]$, 
For a set of bounding boxes $B$, 
TASM picks tile boundaries guided by a desired tile granularity. For coarse-grained tiles (\cref{subfig:coarseGrainTiles}), it places all $B$ within a single, large tile. For fine-grained tiles  (\cref{subfig:fineGrainTiles}), it attempts to isolate non-intersecting $b \in B$ into smaller tiles while respecting minimum tile dimensions specified by the codec and ensuring that no tile boundary intersects any $b \in B$.
\revision{R2D1}{TASM does not limit the number of tiles in a layout. 
% \brandon{Should we mention that codecs often impose an upper-bound?}. 
To modulate quality, this could be made a user-specified setting; we leave this as future work.}
% We evaluate the performance impact of this choice in \cref{subsec:evalNonUniform}
TASM processes fewer pixels from a video stored with fine-grained tiles because the tiles do not contain the parts of the frame between objects, but it processes more individual tiles because multiple tiles in each frame may contain objects. TASM estimates the overall effectiveness of a layout using a cost function that combines these two metrics, as described in \cref{sec:notation}.

In addition to deciding the tile granularity, TASM also chooses \textit{which} objects to design the tile layout around, and therefore which bounding boxes to include in $B$. The best choice depends on the queries. For example, if queries target people, a layout around just people, as in \cref{subfig:customLayoutPed}, is more efficient than a layout around both cars and people (\cref{subfig:customLayoutCarPed}).
We explain how TASM makes this choice in \cref{sec:tilingStrategies}.

\tileLayoutDurationFigure

\subsubsection{Temporally-changing layouts} \label{sec:temporalLayout}

Different tile layouts, uniform and non-uniform, can be used throughout a video; the layout can change as often as every GOP. TASM uses different layouts throughout a video to adapt to objects as they move.
%As described in \cref{sec:videoencoding}, a SOT refers to a sequence of frames with the same tile layout.

% As an example, consider a traffic video that contains cars. The cars move throughout the video, so for a single layout to provide performance improvements, it must provide fast access to every location of every car in the entire duration of the video. This is challenging because as cars move, non-uniform tiles must get larger to contain their trajectory. However, within a small temporal section of the video, the locations of cars are constrained. TASM creates a non-uniform layout that provides fast access to just the subset of car locations in that section.

% The sizes of tiles in uniform layouts are not affected by the layout duration; their dimensions are determined by the number of tiles. However, the sizes of tiles in non-uniform layouts \textit{are} affected by the layout duration.

The size of these temporal sections is determined by the \textit{layout duration}, which refers to the number of frames within a sequence of tiles (SOT). 
Layout duration is separate from GOP length; while layout duration cannot be shorter than a GOP, it can extend over multiple GOPs.
The layout duration affects the sizes of tiles in non-uniform layouts, as shown in  \cref{fig:tileLayoutDurationExample}.
In general, longer tile layout durations have lower storage costs but lead to larger tiles because TASM must consider more object bounding boxes as objects move and new objects appear. Therefore, TASM must decode more data on each frame.
We evaluate this tradeoff in \cref{fig:tileLayoutDurationPlot}.

\subsubsection{Not tiling} \label{sec:notTiling}

%There are cases where tiling is not a useful strategy.

%If the objects targeted by a query cover most of the frame, nearly all of the pixels are part of the query result.

%For example, if an animal gets very close to a nature camera, the animal will take up most of the frame. In this case, tiling cannot provide performance improvements because the target object occupies most of the frame.
Layouts that require TASM to decode a similar number of pixels as when the video is not tiled can actually slow queries down due to the implementation complexities that arise from working with multiple tiles.
Therefore, TASM may opt to not tile GOPs when the gain in performance does not exceed a threshold value. 

% TASM identifies cases where tiling cannot provide improvement based on the number of pixels the layout enables it to skip. As described in \cref{sec:videoencoding}, the decode cost grows with the number of pixels that are decoded. Layouts that require TASM to decode a similar number of pixels as when the video is not tiled can actually slow queries down due to the implementation complexities that arise from working with multiple tiles. TASM sets a threshold for the minimum reduction in decoding cost that a layout must offer to be considered useful.\brandon{I feel like we can cut down on this paragraph a bunch.  Maybe something like `TASM may opt to not tile GOPs when the gain in performance does not exceed a threshold value.'}
% This is similar to how indices can sometimes slow down queries in relational databases.

\subsubsection{Data storage and retrieval} \label{sec:files}
% \todo{discuss how tiled videos can be made accessible to other software that does not accept tiles}
% \todo{mention that this can work with blob storage systems because only the portions of the video that are read can be saved as tiles}

% \tileAndFilesFigure
% For example if a video is encoded with a tile layout of three rows and three columns, TASM stores nine video files, each of which can be decoded individually. Each video acts as a spatial random access point for the region of the frame it contains.

TASM stores each tile as a separate video file, as shown in \cref{fig:tileAndFiles}. If different layouts are used throughout the video, each tile video contains only the frames with that layout. 
%  using a separate video for every tile. 
If only a segment of a video is ever queried, TASM reads and tiles just the frames in that segment.
\revision{R3D6}{This storage structure facilitates the ingestion of new videos because each video's data is stored separately. Additionally, because each GOP is also stored separately, to modify an existing video, updated GOPs
  can replace original ones, or new GOPs can be appended.}
% with the updated ones.}

% In \Cref{fig:tileAndFiles}, all of the tiles for frames $[j+1, n]$ are stored in the directory {\tt video/frames\_j+1-n/} and the upper-left portion of these frames is saved in the video {\tt video/frames\_j+1-n/tile0.mp4}\brandon{This can be compressed a bunch}.

%\subsection{Tile-based data access}
%\label{sec:query}

%TASM quickly accesses query regions by utilizing the per-tile videos described in \cref{sec:files}, which act as spatial random access points. When a query is processed, TASM only decodes the tile videos that contain the regions specified by the query, as shown in \cref{fig:queryToTiles}.
TASM retrieves just the tiles containing the objects targeted by queries.
% For a query of an object contained within a single tile, TASM retrieves only the relevant tile. For queries over objects spanning multiple tiles, TASM must retrieve and combine the contents of multiple tiles\brandon{Compress and combine previous two sentences into one}.
When complete frames are requested, TASM applies homomorphic stitching (see \cref{sec:videoencoding}).
% When a query asks for the entire frame, TASM uses homomorphic stitching to recover the frame as described in \cref{sec:videoencoding}\brandon{Kill this sentence and say something like `When complete frames are requested, TASM applies homomorphic stitching (see Section 2).'}.
This stitching process can also be used to efficiently convert the tiles into a codec-compliant video that other applications can interact with.
% Because the tiles are stored as standard videos, the system can decode the tiles just like it would any video.

% As a concrete example, consider a query that retrieves the pedestrians in a crosswalk. During query execution, the query processor requests from TASM the video region containing the crosswalk, and TASM retrieves this by decoding only the tiles that contain part of the crosswalk.

%!TEX root = paper.tex

\section{Tiling strategies} \label{sec:tilingStrategies}

% The previous section describe at a high level various strategies TASM can use to partition a video into tiles. This section describes how TASM navigates tradeoffs between the various strategies to pick

\squeeze{TASM automatically tunes the tile layout of a video to improve query performance. The objects in a video
% may be \textit{known} or \textit{unknown}. Similarly,
and workloads, or the set of queries presented to a VDBMS, may be \textit{known} or \textit{unknown}. When TASM has full knowledge of both the objects targeted by queries and the locations of these objects in video frames, TASM designs tile layouts before queries are processed, as described in \cref{sec:knownWorkload}. In practice, the objects targeted by queries and their locations are initially unknown. TASM uses techniques from online indexing to incrementally design layouts based on prior queries and the objects detected so far, as described in \cref{sec:unknownWorkloadUnknownObjects}. Finally,
  TASM also creates an efficient, initial tiling before any queries are executed as we present in \cref{sec:ROI}.
}

%More commonly, the locations of objects are initially unknown and must be incrementally detected as queries are processed. When the types of objects targeted by queries are known, TASM designs tile layouts around these objects as they are detected, as described in \cref{sec:knownWorkloadUnknownObjects}. However, 

% More commonly, TASM will know which objects will be queried, but not where they are. \magda{The description of this case sounds identical to the next case. In both cases, we must incrementally find the objects.} In this scenario, TASM designs tile layouts incrementally as objects are detected, as described in \cref{sec:knownWorkloadUnknownObjects}. Finally, \cref{sec:unknownWorkloadUnknownObjects} describes how, when both the objects targeted by queries and their locations are unknown, TASM uses techniques from online indexing to incrementally design tile layouts based on previous queries and the objects that have been detected so far. 

% \dong{As described in \cref{sec:edgeTiling}}, the processing performed by TASM can be augmented by utilizing the compute power of edge cameras that originally capture the video. Edge cameras can produce semantic information at the time of video capture and use this metadata to initialize the semantic index and encode the video with tiles even before it reaches a VDBMS \dong{, which avoids the re-encoding cost for tiling after videos were encoded initially without tiles at the edge}.

\vspace{-0.5em}
\subsection{Notation and cost function} \label{sec:notation}

We first introduce notation that will be used throughout this section. A query workload $Q=(q_1, ..., q_n)$ is a list of queries, where each query requests pixels belonging to specified object classes, possibly with temporal constraints.
%  on the frames the objects must appear in.
 The set $O_{q_i}$ represents the objects requested by an individual query $q_i$, while $O_Q=\cup_{q_i \in Q} O_{q_i}$ is the set of all objects targeted by
  % queries in
$Q$.
% , and $O^*_v$ is the set of all objects in the semantic index for video $v$.

A video $v = s_0 \oplus \cdots \oplus s_n$ is a series of concatenated, non-overlapping, non-empty sequence of tiles (SOTs; see \cref{sec:videoencoding}), $s_i$. A video layout specification $\LayoutSpec {=} s_i \mapsto L$ maps each SOT to a tile layout, $L$, which specifies how frames are partitioned into tiles, as described in \cref{sec:videoencoding}. If a SOT is not tiled, then $s_i {\mapsto} \omega$, where $\omega$ refers to a $1 {\times} 1$ tile layout. \Partition{s}{O} refers to tiling the SOT using a non-uniform layout around the bounding boxes associated with objects in the set $O$ using the techniques from \cref{sec:nonUniformLayouts}. For example, \Partition{s}{\{car, person\}} refers to creating a layout around cars and people, as in \cref{subfig:customLayoutCarPed}.
%  while \Partition{s}{\{car\}} refers to creating a layout around just cars, as in \cref{subfig:customLayoutCar}.

TASM 
% uses a cost function to 
implements a ``what-if'' interface~\cite{DBLP:conf/sigmod/ChaudhuriN98} to estimate the cost of executing queries with alternative layouts using a cost function.
The estimated cost of executing query $q$ over SOT $s$ encoded with layout $L$ is $C(s, q, L) {=} \beta \cdot P(s, q, L) + \gamma \cdot T(s, q, L)$.  The cost $C$ is proportional to the number of pixels $P$, and the number of tiles $T$ that are decoded, both of which depend on the query and layout. 
To validate this cost function and estimate $\beta$ and $\gamma$ to use in experiments, we fit a linear model to the decode times for over $1{,}400$ video, query, and non-uniform layout combinations used in the microbenchmarks in \cref{sec:microbenchmarks}. The resulting model achieves $R^2 {=} 0.996$.
The exact values of $\beta$ and $\gamma$ will depend on the system; \squeeze{TASM can re-estimate them by generating a number of layouts from a small sample of videos and measuring execution time.}

Finally, the cost of executing $q$ over video $v$ encoded with layout specification \LayoutSpec is the sum of its SOT costs (i.e., $C(v, q, \LayoutSpec) {=} \sum_{s_i \in v} C(s_i, q, \LayoutSpec(s_i))$) and
the cost of executing an entire query workload is the sum over all individual queries, $C(v, Q, \LayoutSpec) {=} \sum_{q_i \in Q} C(v, q_i, \LayoutSpec)$. 
% The cost of executing a query over a single SOT $g$ with layout $L$ is $C(g, q_i, L)$, which measures only the pixels and tiles decoded in the specified SOT.
% Once values for $\beta$ and $\gamma$ are determined, TASM can predict the performance of tile layouts.
% The improvement in query time on query $q$ over the entire video achieved with layout specification $\LayoutSpec'$ compared to $\LayoutSpec$ is $\Delta(q, \LayoutSpec, \LayoutSpec' ) = C(v, q, \LayoutSpec) - C(v, q, \LayoutSpec')$. 
The difference in estimated query time for query $q$ over SOT $s$ between layouts $L$ and $L'$ is $\Delta(q, L, L', s) {=} C(s, q, L) {-} C(s, q, L')$, or simply $\Delta(q, L, L')$ when $s$ is obvious from the context. The cost of (re-)encoding SOT $s$ with layout $L$ is $R(s, L)$. 
% The cost of re-encoding a video with layout specification \LayoutSpec is $R(v, \LayoutSpec) = \sum_{g_i \in v} R(g_i, \LayoutSpec(g_i))$.

\revision{R2W1}{
Using this cost function, the maximum expected improvement for an individual query is inversely proportional to the object density, which determines the number of pixels ($P$) and tiles ($T$). Tiling therefore leads to negligible improvement---or even regressions---when objects are dense and occupy a large fraction of a frame. In those cases, TASM does not tile a video at all as we discuss in \cref{sec:knownWorkload}. In contrast, tiling yields large improvements when objects are sparse. \Cref{fig:tilingCutoff} shows the linear relationship. It shows how, for a given video and query, non-uniform tiling reduces the number of pixels that must be decoded, which directly increases performance. TASM's regret-based approach described in \cref{sec:unknownWorkloadUnknownObjects} converges to such good layouts over time as queries are executed. \Cref{fig:tileGranularityPlot} also shows how object densities affect performance.
% Using this cost function, the maximum expected improvement for a query is inversely proportional to the object density, which scales $P$ and $T$.
% We therefore expect negligible improvement or even regressions when objects are dense and occupy many pixels and/or tiles, and large improvements when objects are sparse.
% \Cref{subfig:same} shows this correlation: when objects are sparse and occupy $<20\%$ of each frame, decode costs improve by around $80\%$ compared to the untiled video. However, when objects are dense the improvement can be much less. 
% The incremental approach described in \cref{sec:unknownWorkloadUnknownObjects} allows TASM to eventually find the layouts that achieve these savings, if possible.
}

\subsection{Known queries and known objects} \label{sec:knownWorkload}

We first present TASM's fundamental video layout optimization assuming a known workload, meaning that TASM knows which objects will be queried,
% in each part of the video
\textit{and} the semantic index contains their locations.
These assumptions are unlikely to hold in practice, and we relax them in the next section.

Given a workload and a complete semantic index, TASM decides on SOT boundaries then picks a tile layout for each SOT to minimize execution costs over the entire workload. More formally, the goal is to partition a video into SOTs, $v = s_0 \oplus \cdots \oplus s_n$ and find $\LayoutSpec^* = \mathrm{arg\,min}_{\LayoutSpec} C(v, Q, \LayoutSpec)$.

% \magda{We should say here that the experiment in Figure 10 motivates us to use one SOT for one GOP. Currently, there is no justification for doing so. Or, we can say here that the number of GOPs that go into one SOT is a tunable parameter. And, we find in Figure 10 that using one SOT equal to one GOP gives the best performance for less than 10\% of space overhead, so this is the approach that we use in our prototype system.}
% \brandon{I feel like we should say something like `To minimize the optmization function, TASM must make two decisions.  First, deciding SOT boundaries.  Second, choosing a layout for each SOT.' And then the next two paragraphs address each decision.  But, space...}
The experiment in \cref{fig:tileLayoutDurationPlot} motivates us to create small SOTs because %it leads to the best performance. 
they perform best. We therefore partition the video
such that each GOP %in the original video
corresponds to a SOT in the tiled video.
% into SOTs at GOP boundaries because it has the best performance, so each GOP in the original video corresponds to a SOT in the tiled video. 
This produces a tiled video with a similar storage cost as the untiled video because it has the same number of keyframes.
% As mentioned in \cref{sec:temporalLayout}, shorter layout durations lead to greater performance improvements, though they also incur higher storage costs\brandon{Does TASM support exploiting this tradeoff?  If not, why are we making this point?}. This tradeoff is shown in \cref{fig:tileLayoutDurationPlot}.

% The set of all possible layouts for a given SOT is extremely large; it contains all possible uniform and non-uniform layouts. It would be too expensive for TASM to evaluate every possible layout.
It would be too expensive for TASM to consider every possible layout, uniform and non-uniform, for a given SOT.
However, tile layouts that isolate the queried objects should improve performance the most. % lead to the greatest performance improvements.
Additionally, we empirically demonstrate that non-uniform layouts outperform uniform layouts (see \cref{subfig:tilingOnQueryTime}), and that fine-grained layouts outperform coarse-grained layouts (see \cref{fig:tileGranularityPlot}).
% Additionally, we empirically demonstrate that fine-grained non-uniform layouts outperform coarse-grained and uniform layouts, as shown in \cref{subfig:tilingOnQueryTime} and \cref{fig:tileGranularityPlot}.
Therefore, for each $s_i$, TASM only considers a fine-grained, non-uniform layout around the objects targeted by queries in that SOT, $O_{s_i} \subseteq O_Q$.
% We call this the ``known-query/known-object'' (KQKO) optimization. 

TASM's optimization process proceeds in two steps. First, for each $s_i$ and associated layout, $L {=} \Partition{s_i}{O_{s_i}}$, TASM estimates if re-tiling the SOT with $L$ will improve query performance at all. As described in \cref{sec:notTiling}, TASM does not tile $s_i$ when $P(s_i, Q, L) {>} \alpha {\cdot} P(s_i, Q, \omega)$, where $\alpha$ specifies how much a tile layout must reduce the amount of decoding work compared to an untiled video (i.e., $L{=}\omega$). In our experiments we find $\alpha {=} 0.8$ to be a good threshold. As shown in \cref{fig:tilingCutoff}, this value of $\alpha $ prevents TASM from picking tile layouts that would slow down query processing, but does not cause it to ignore layouts that would have significantly sped up queries. Second, from among all such layouts, TASM selects the layout with the smallest estimated cost for the workload.

\vspace{2em}
\subsection{Unknown queries and unknown objects} \label{sec:unknownWorkloadUnknownObjects}

\incrementalLayoutPseudocode

In practice, objects targeted by queries and their locations are initially unknown. Physically tuning the tile layout is then similar to the online index selection problem in relational databases~\cite{DBLP:conf/icde/BrunoC07}. In both, the system reorganizes physical data or builds indices with the goal of accelerating unknown future queries.
However, while a nonclustered index can benefit queries over relational data because there are many natural random access points, video data requires physical reorganization to introduce useful random access opportunities.
% \magda{If the problem is similar, then how does our solution compare? We need to explain briefly here.}
As TASM observes queries and learns the locations of objects, it makes incremental changes to the video's layout specification to introduce these random access points.

TASM optimizes the layout of each SOT independently because
each SOT's contribution to query time and the cost to re-encode it are independent of other SOTs.
% This allows TASM to optimize the layout of each SOT independently.
TASM optimizes the layout of an SOT based on the queries that have targeted it so far.
TASM may even tile it multiple times with different layouts as the semantic index gains more complete information and TASM observes queries that target additional objects.
%Whereas in \cref{sec:knownWorkloadUnknownObjects} TASM had complete knowledge of $O_Q$ and therefore no uncertainty about whether it should re-tile a SOT or wait for more objects to be detected, in this scenario TASM must deal with uncertainty when creating layouts because it does not know whether it has seen all of $O_Q$, or whether future queries will target new objects.
% \magda{I feel like between the paragraph above and this one, we repeats multiple times that we optimize each SOT layout separately. We should remove this redundancy.}

% the video transitions through a series of layout specifications $\mathbf{\mathcal{L}} = [\LayoutSpec_0, \cdots, \LayoutSpec_n]$. 
% After observing query $q_i$, TASM picks $\LayoutSpec_{i+1}$ and re-tiles the SOTs of the video whose layouts changed, $\{ g \in v \mid \LayoutSpec_i(g) \neq \LayoutSpec_{i+1}(g) \}$, with a total re-tiling cost of $R(v, \LayoutSpec_{i+1})$. The improvement in query time on $q_{i+1}$ achieved by re-tiling $v$ with $\LayoutSpec_{i+1}$ is $\Delta(q_{i+1}, \LayoutSpec_i, \LayoutSpec_{i+1})$.
%  For an arbitrary SOT $s_j$, $\mathbf{L} = [L_0^j, \cdots, L_n^j ]$ denotes the sequence of layouts it transitions through over the workload.

As TASM re-encodes portions of the video,
the SOT $s_j$ transitions through a series of layouts: $\mathbf{L} {=} [L_0^j, \cdots, L_n^j ]$.
TASM's goal is to pick a sequence of layouts that minimizes the total execution cost over the workload by finding $\mathbf{L}^* {=} \mathrm{arg\,min}_{\mathbf{L}}\sum_{q_i \in Q} ( C(s_j, q_i, L^j_i) + R(s_j, L^j_{i}) )$.
The first term measures the cost of executing the query with the current layout, and the second term measures the cost of transitioning the SOT to that layout. If the layout does not change (i.e., $L^j_{i-1}{=}L^j_i$), then $R(s_j, L^j_{i}){=}0$.
However, future queries are unknown, so TASM must pick $L^j_{i+1}$ without knowing $q_{i+1}$. Therefore, TASM uses heuristics to pick a sequence of layouts, $\hat{\mathbf{L}}$, that approximates $\mathbf{L}^*$.
While there are no guarantees on how close $\hat{\mathbf{L}}$ is to $\mathbf{L}^*$, we show in \cref{sec:workloads} that empirically these layouts perform well.
% which is hopefully \magda{``hopefully'' is word that should not appear in a technical paper. Is it or is it not as good? Are there some known guarantees? Yes or no. If no, we can say tha twe should experimentally.} nearly as good as \magda{``nearly as good as'' is very vague. How good is ``nearly as good as''?} $\mathbf{L}^*$.
% TASM does not know which objects will be targeted by future queries\brandon{I think you already said this above.}, but because many applications query for similar objects over time\brandon{Is this a heuristic captured by $\hat{L}$?}
One such heuristic is guided by the observation that 
many applications query for similar objects over time. TASM therefore creates layouts optimized for objects it has seen so far. More formally, let $O_{Q'}$ be the set of objects from $Q' {=} (q_0, \cdots, q_i) \subseteq Q$.
TASM only considers non-uniform layouts around objects in $O_{Q'}$ for $L_{i+1}$.
% When TASM considers layouts for $L_{i+1}$, it only considers non-uniform layouts around objects in $O_{Q'}$\brandon{I feel like this paragraph could be pared down and pushed into the previous.}. 
% TASM updates $O_{Q'} \gets O_{Q'} \cup O_{q_{i}}$ after executing each query $q_{i}$.

Now consider a future query $q_j$ that targets a new class of object: $O_{q_j} {\not \subseteq} O_{Q'}$. While $L_{i+1}$ will not be optimized for $O_{q_j}$, TASM attempts to create layouts that will not hurt the performance of queries for new types of objects. It does this by creating fine-grained tile layouts because, as shown in \cref{fig:tileGranularityPlot},
fine-grained tiles lead to better query performance than coarse-grained tiles when queries target 
% objects that were not considered when creating the layout 
new types of objects
($\Partition{s}{O'}, O' {\cap} O_{q_j} {=} \emptyset$).
% when queries target objects that were not considered when creating the tile layout ($\Partition{s}{O'}, O' \cap O_{q_j} {=} \emptyset$), fine-grained tiles lead to better query performance than coarse-grained tiles. 
Objects that are not considered when designing the tile layout may intersect multiple tiles, and it is more efficient for TASM to decode all intersecting tiles when the tiles are small, as in fine-grained layouts, than when the tiles are large, as in coarse-grained layouts.

At a high level, TASM tracks alternative layouts based on the objects targeted by past queries and identifies potentially good layouts from this set by estimating their performance on observed queries.
% \magda{This is the first time that we introduce the term ``regret''. We should pause and say that we build on regret-minimization techniques. We need to justify why we made that choice. And we should cite the related work that uses regret minimization.}
TASM's incremental tiling algorithm builds on related regret-minimization techniques~\cite{DBLP:conf/icde/DashKA09, DBLP:conf/icde/BrunoC07}. Regret captures the potential utility of alternative indices or layouts over the observed query history when future queries are unknown.
As each query executes, TASM accumulates regret $\delta_k^j$ for each SOT $s_j$ and alternative layout $L_k$, which measures the total estimated performance improvement compared to the current tile layout over the query history.

% The regret $\delta_k^j$ for each $s_j$ and alternative layout $L_k$ measures the total estimated performance improvement \magda{We need to specify the improvement of what compared to what.} over the query history.

\Cref{alg:incrementalLayout} shows the pseudocode of our core algorithm for incremental tile layout optimization using regret minimization.
Initially, TASM has not seen queries for any objects, so it does not have any alternative layouts to consider, and each SOT is untiled (line 1).
% \magda{We need to say: ``Figure 6 shows the pseudocode of our core algorithm for incremental tile layout optimization using regret minimization.'' Only after that we can refer to line numbers in the algorithm.}
After each query, TASM updates the set of seen objects and alternative layouts (lines 3-4). Each potential layout is a subset of the seen objects that have location information in the semantic index.
TASM then accumulates regret for each potential layout by computing $\Delta$ and adding it to $\delta$.
$\Delta$ measures the estimated performance improvement of executing the query with an alternative layout rather than the current layout, using the cost function described in \cref{sec:notation}: $\Delta(q, L, L') = C(s, q, L) - C(s, q, L')$.
Layouts with high $\Delta$ values would likely reduce query costs, while layouts with low or negative values could hurt query performance. TASM accumulates these per-query $\Delta$'s into regret to estimate which layouts would benefit the entire query workload.
% The regret is updated \magda{Aha! Move this up first and say specifically that $\Delta$ measures the incremental regret. And that we ``accumulate regret'' by computing the $\Delta$ and adding it.} with the term $\Delta$ which measures the estimated performance improvement of executing the query with an alternative layout rather than its current layout, using the cost function described in \cref{sec:notation}. \magda{This is really important. We should show the exact equation for $\Delta$, show how it uses the cost function. This is the core of the algorithm, we should show the details and discuss the details here in the text.}

TASM first retroactively accumulates regret for new layouts based on the previous queries (lines 5-7), and then accumulates regret for the current query (lines 9-10). 
Finally, TASM weighs the performance improvements against the estimated cost of transitioning a SOT to a new layout. In lines 11-14, TASM only re-tiles $s_j$ once its regret exceeds some proportion of its estimated retiling cost: $\delta_k^j > \eta \cdot R(s_j, L_k)$.
% \magda{Here too, I would put the equation. Assume that the reader may not look at the pseudocode. We should repeat all key equations here in the text.}

% % % <original text>
% After executing each query, TASM must decide whether to update the layout of each SOT $s_j {\in} v$. TASM maintains a set of alternative layouts, $L_{alt} {=} \{ L_0, \cdots, L_m \}$, where each potential layout partitions around a subset of the seen objects that have location information in the semantic index, $\Partition{s_j}{O'}, O' \subseteq O_{Q'}$. TASM identifies potentially good layouts by estimating the performance improvements that each alternative layout could have provided on queries in $Q'$.

% As each query executes, TASM accumulates regret~\cite{DBLP:conf/icde/DashKA09} $\delta_k^j$ for each $s_j$ and alternative layout $L_k$, which measures the total estimated performance improvement over the query history.
% After each $q_i$, TASM estimates $\forall s_j {\in} v, L_k {\in} L_{alt}, \delta_k^j {=} \delta_k^j + \Delta(q_i, L^j_i, L_k)$, where initially each $\delta^j_k {=} 0$ when the first query is executed.
% $\Delta(q_i, L^j_i, L_k)$ measures the estimated performance improvement of executing the query on $s_j$ with an alternative layout rather than its current layout, $L^j_i$, using the cost function described in \cref{sec:notation}.
% % % <\original text>

\revision{R1D3}{As an example, consider a city planning application looking through traffic videos for instances where both cars and pedestrians were in the crosswalk at the same time.} 
% \magda{If that's the application, why is the second query asking about people?} \maureen{Good point. I thought it may be creepy to describe a query that tries to identify a person, so I switched to a different application.}
% As an example, consider the amber alert application from \cref{sec:introduction}.
Initially the traffic video is untiled, so for each $s_i$, $\LayoutSpec(s_i) {=} \omega$.
%  $\forall s_i, \LayoutSpec(s_i) {=} \omega$.
% Initially $L_{alt}$ is empty.
Suppose the first query requests cars in $s_0$.
TASM updates $L_{alt} {=} \{\{car\}\}$ to consider layouts around cars.
TASM accumulates regret for $s_0$ as $\delta^0_{car} {=} \Delta(q_0, \omega, \Partition{s_0}{\{car\}})$, and it is positive because tiling around cars would accelerate the query.
Suppose the next query is for people in $s_0$.
TASM updates $L_{alt} {=} \{ \{car\}, \{person\}, \{car, person\} \}$ to consider layouts around cars and people.
The regret for \Partition{s_0}{\{car\}} on $q_1$ will likely be negative because layouts around anything other than the query object tend to perform poorly (see \cref{subfig:different}), so $\delta^0_{car}$ decreases.
TASM retroactively accumulates regret for the new layouts.
The accumulated regret for $\Partition{s_0}{\{person\}}$ will be similar to $\delta^0_{car}$ because it would accelerate $q_1$ and hurt $q_0$. 
\Partition{s_0}{\{car, person\}} has positive regret for both $q_0$ and $q_1$, so after both queries it has the largest accumulated regret.

% In addition to considering the performance improvements offered by alternative layouts, TASM must consider the cost of transitioning $s_j$ to a new layout; it estimates the cost of $R(s_j, L_k)$ based on the encoding performance of the system. 
% TASM re-tiles $s_j$ with $L_k$ when $\delta^j_k > \eta \cdot R(s_j, L_k)$. 
The threshold $\eta$ (see line 13) determines how quickly TASM re-tiles the video after observing queries for different objects.
Using $\eta=0$ risks wasting resources to re-tile SOTs. The work to re-tile could be wasted if a SOT is never queried again because no queries will experience improved performance from the tiled layout. The work to re-tile can also be wasted if queries target different objects because TASM will re-tile after each query with layouts optimized for just that query. Values of $\eta > 0$ enable TASM to observe multiple queries before picking layouts, so the layouts can be optimized for multiple types of objects. Observing multiple queries before committing to re-tiling also enables TASM to avoid creating layouts optimized for objects that are infrequently queried because layouts around more representative objects will accumulate more regret. However, if the value of $\eta$ is too large, it reduces the number of queries whose performance benefits from the tiled layout. Using a value of $\eta=1$ is similar to the logic used in the online indexing algorithm in \cite{DBLP:conf/icde/BrunoC07}, and we find it generally works well in this scenario, as shown in ~\cref{fig:visualroadWorkloads}.
If the types of objects queries target changes, this incremental algorithm will take some amount of time to adjust to the new query distribution, depending on the value of $\eta$.

% \dong{This subsection (C) seems long. It could be cut if running out of space.}

%\subsection{Predicting tile layout performance} \label{sec:predictPerformance}

%TASM attempts to pick tile layouts that will improve query performance, but it cannot simply re-encode the video with every candidate layout to get empirical performance results because that would be too computationally expensive. Instead, it must estimate the performance of layouts using metrics that can be cheaply computed. As described in \cref{sec:notation}, query execution time is correlated with the number of pixels and number of tiles that must be processed by the decoder: $C(v, q_i, \LayoutSpec) = \beta \cdot P(v, q_i, \LayoutSpec) + \gamma \cdot T(v, q_i, \LayoutSpec)$. Given a query and layout, these are both metrics that can be computed without actually encoding the video based on tile/bounding box intersections and keyframe locations.

%We fit a linear model to the decode times for over $1,400$ video, query object, and non-uniform layout combinations to find $\beta$ and $\gamma$, and the resulting model achieves $R^2 = 0.996$. TASM uses these values to implement a ``what-if'' interface~\cite{DBLP:conf/sigmod/ChaudhuriN98} that estimates the cost of executing queries with alternate layouts.

\subsection{ROI tiling}\label{sec:ROI}

% \magda{This introduction is too abrupt. I recommend starting by saying that initially, nothing is known about a video. As we discussed in the introduction, the first object detection query will do a full scan phase and will apply a simple predicate. We can accelerate this full scan through predicate pushdown. TASM does this by identifying regions of interest, because there is often no point applying the full-scan predicate to all parts of a frame. Then we should say directly that the detection of regions of interest is a user-specified predicate and enumerate the various options.}

Initially, nothing is known about a video.
As we discussed in \cref{sec:introduction}, in many systems, the first object detection query performs a full scan and applies a simple predicate to filter away uninteresting frames or compute statistics.
Because of the speed of these initial filters, decoding and preprocessing is the bottleneck for this phase~\cite{DBLP:journals/corr/abs-2007-13005}.
To accelerate this full scan phase, TASM also uses predicate pushdown. Instead of creating tiles around objects,
however,  TASM creates tiles around more general \textit{regions of interest} (ROIs), where objects are expected to be located.
ROIs are defined by bounding boxes, so TASM uses the same tiling strategies described in previous sections.
TASM accepts a user-defined predicate that detects ROIs and inserts the associated bounding boxes into TASM's semantic index.
Examples include applying background subtraction to identify foreground objects, running specialized models trained to identify a specific object type~\cite{DBLP:conf/sigmod/LuCKC18,DBLP:journals/pvldb/KangEABZ17}, extracting motion vectors to isolate areas with moving objects, or any other inexpensive computation.
More expensive predicates may also be used by applying them every $n$ frames, as in \cite{DBLP:conf/sigmod/BastaniHBGABCKM20}.

\revision{R1D4}{
Generating ROIs and creating tiles around these regions are operations that a compute-enabled camera can perform directly as it first encodes the video.
Cameras are now capable of running these lightweight predicates as video is captured~\cite{boschCameraTrainer}. For example, specialized background subtractor modules can run at over 20 FPS on low-end hardware~\cite{fastestBackgroundSubtraction}. This optimization is designed to be implemented on the edge.
}

Through its semantic predicate pushdown optimization, TASM improves the performance of object detection queries by only decoding tiles that contain ROIs.
As we show in \cref{sec:objectDetectionEval}, an initial ROI layout in combination with semantic predicate pushdown can significantly accelerate the full scan phase of object detection queries while maintaining accuracy.
%!TEX root = paper.tex

\section{Evaluation} \label{sec:evaluation}

\videoSetsTable

We implemented a prototype of TASM in C++ integrated with LightDB~\cite{LightDB:pvldb/HaynesMABCC18}.
TASM encodes and decodes videos using NVENCODE/NVDECODE~\cite{nvenc} with the HEVC codec. We perform experiments on a single node running Ubuntu 16.04 with an Intel i7-6800K processor and \squeeze{an Nvidia P5000 GPU}. Our prototype does not parallelize encoding or decoding multiple tiles at once. We use FFmpeg~\cite{ffmpeg} to measure video quality.

We evaluate TASM on both real and synthetic videos with a variety of resolutions and contents as shown in \Cref{table:videoSets}.
Visual Road videos simulate traffic cameras. \revisioncolor{They include stationary videos as well as videos taken from a roof-mounted camera (the latter created using a modified Visual Road generator~\cite{DBLP:conf/sigmod/HaynesMBCC19})} The Netflix datasets primarily show scenes of people. The XIPH dataset captures scenes ranging from a football game to a kayaker. The MOT16 dataset contains busy city scenes with many people and cars. The El Fuente video contains a variety of scenes (city squares, crowds dancing, car traffic). 
In addition to evaluating the full El Fuente video, we also manually decompose it into individual scenes using the scene boundaries specified in \cite{cdvlElFuente} and evaluate each independently.
% \magda{Here, we need to give one sentence for each entry in Table 1.}
% For example, Visual Road videos simulate traffic cameras, while scenes from the MOT16 and El Fuente capture busy scenes with many objects. \magda{``Busy scenes with many objects'' is too vague. Can we say something more specific like: ``Scenes of crowds dancing, car traffic, and similar''?}
We do not evaluate on videos with resolution below 2K because we found that decoding low-resolution video did not exhibit significant overhead. 
All experiments populate the semantic index with object detections from YOLOv3~\cite{DBLP:journals/corr/abs-1804-02767}, except for the MOT16 videos where we use the detections from the dataset~\cite{DBLP:journals/corr/MilanL0RS16}. \squeeze{We store the semantic index using SQLite~\cite{sqlite},
    % We use SQLite~\cite{sqlite} to store semantically indexed data, 
and TASM maps bounding boxes to tiles at query time.}
%  though a future enhancement involves pre-computing and storing this mapping.}

The queries used in the microbenchmarks evaluated in \cref{sec:unifVsCustom} and \ref{sec:microbenchmarks} are \revision{R3D3}{subframe selection} queries of the form ``\texttt{\small SELECT o FROM v}'', which cause TASM to decode all pixels belonging to object class \texttt{o} in video \texttt{v}. The queries used in the workloads in \cref{sec:workloads}
additionally include a temporal predicate (i.e.,
``\texttt{\small SELECT o FROM v WHERE start < t < end}''). 
\footnote{\revision{R3D7}{While we use SQL to explain the experiments because of its familiarity to most readers, other language bindings on TASM's API are possible; the language itself is not the focus of this paper.
}}
\squeezemore{Reported query times include both the index look-up time and the time to read from disk and decode the tiles. }

Unless otherwise specified, queries target the most frequently occurring objects in each video. When videos primarily show a single type of object (e.g., some Netflix public dataset videos show only people), queries target just that object. When videos feature multiple types of objects with similar frequency (e.g., the Visual Road videos show similar numbers of cars and people), we evaluate on queries that target each object type.
Queries over the MOT16 videos retrieve cars \textit{and} people because the bounding boxes that come with the dataset are unlabeled, so we store them in the semantic index with a generic label of ``object''.
For all graphs, the bars show the median value \revision{R1D2}{across videos}, while the error bars denote the interquartile range (IQR) \revision{R1D2}{across videos. The performance differs across videos because they have different object densities, which affects TASM's efficacy as described in \cref{subsec:evalNonUniform}.
However, the runtime for a single query on any video has low variance. The standard deviation for multiple executions of the same query is $<1\%$ of that query's mean execution time.
}
% \todo{R1D2: Clarify that bars are across videos; add some mention of the scale of intra-video variance.}

% \magda{This last paragraph is very confusing. Especially that Figure 8 also shows results for querying the wrong object. Maybe explain which objects are queried in each subsection. Or we can say that: ``Unless otherwise specified, queries ask for the most frequently occurring object...'' And then later specify otherwise as needed. We should also clarify that ``person'' is not the most frequent object in all the videos but ``car'' is sometimes more frequent.. Otherwise, it sounds like it's possible we queried only for people in all the videos.}
% For each input video, queries target the object classes that occurred most frequently in the video. Some videos feature a single object class that occurs much more frequently than any other. For example, some videos in the Netflix public dataset~\cite{netflixPublicDataset} only show people. However, if multiple object classes have similarly large numbers of detections, we evaluate the microbenchmarks on each object class (e.g., cars and people in the Visual Road~\cite{DBLP:conf/sigmod/HaynesMBCC19} traffic videos). We also use these different object classes in \cref{sec:workloads} to evaluate workloads with queries for multiple types of objects.

\subsection{Tiling effect on decode cost and quality} \label{sec:unifVsCustom}
\tileImprovementsPlot

% \magda{This section should be shortened. The result is clear: Non-uniform layouts yield better query performance and better video quality. If we need space, we should condense this description.}

We first evaluate whether tiling can provide meaningful improvements in query time without degrading the visual quality of videos. 
We find that non-uniform layouts yield better query performance and higher video quality than uniform layouts.
% \Cref{subfig:tilingOnQueryTime} shows the improvement in query time achieved by tiling videos compared to executing queries over a video that is not tiled. 
\Cref{fig:tileImprovementsPlot} only shows results for videos and queries that benefit from tiling, using the layouts that empirically led to the greatest performance improvement. We discuss how TASM determines whether to tile a video in \cref{sec:notTilingEval} and how it selects the optimal tile layout in \cref{sec:microbenchmarks} and \cref{sec:workloads}.

\Cref{subfig:tilingOnQueryTime} shows the improvement in query time achieved by operating over a tiled video compared to a video that is not tiled.
%  that tiling a video speeds up queries, and non-uniform layouts lead to greater performance improvements than uniform layouts.
% Over all videos and query objects, the best uniform layout gives an average of 37\% improvement in decode time, and the best non-uniform layout gives an average of 51\% improvement.
For a given video and query object, a non-uniform layout provides an average of 10\% improvement and up to a 35\% improvement over the best uniform layout.

\Cref{subfig:tilingOnQuality} shows that tiling maintains good visual quality when the tiles are stitched to recover the full frame. 
We measure quality using peak signal-to-noise ratio (PSNR), where 
% We use peak signal-to-noise ratio (PSNR) to measure quality. PSNR
values above 30 dB are acceptable~\cite{DBLP:conf/asplos/LottariniRCKRSW18}, and videos with values $\geq 40$ dB are perceived to have good quality~\cite{vranjes2008locally, netflixPublicDataset}. %videoClarity
PSNR was computed over the entire tiled video stitched using homomorphic stitching~\cite{LightDB:pvldb/HaynesMABCC18} and compared against the untiled video. 
For comparison, the median PSNR after re-encoding the videos without tiles is 46 dB.
Non-uniform layouts achieve an average PSNR of 40 DB, while uniform layouts have an average of 36 dB.
% Videos tiled with the non-uniform layouts have an average PSNR of 40 dB, but that drops to 36 dB when they are tiled with the uniform layouts.
% Videos tiled with the uniform layout have an average PSNR of 36 dB, and videos tiled with the best non-uniform layout have an average PSNR of 40 dB.
PSNR is likely lower for the uniform layouts because the layouts with the largest performance improvement have many tiles (the median number of tiles is 25), and therefore a large number of tile boundaries where quality is degraded.

\subsection{Microbenchmarks} \label[section]{sec:microbenchmarks}

\subsubsection{Uniform tiles} \label{subsec:uniformtileseval}
\numUniformTilesPlot

% I removed this since we say it above already.
% only includes data for videos and objects where at least one uniform layout improved query performance. $30\%$ of the inputs were not improved by tiling and were omitted from this experiment; in \cref{sec:notTilingEval} we show how TASM identifies these inputs. 
%\Cref{subfig:tilingOnQueryTime}

We dig deeper into the results of \Cref{fig:tileImprovementsPlot} and show in \Cref{fig:numUniformTilesOnQueryTime} the performance improvements when varying the number of uniform tiles on the same set of videos.
\revision{R3W3, R3D5}{We increase the number of uniform tiles first by increasing the number of rows and columns together, and then by only increasing the number of columns once the height of each tile reached the minimum height allowed by the decoder.}
\Cref{fig:numUniformTilesOnQueryTime} shows that creating more uniform tiles initially improves query time because tiles contain fewer non-object pixels.
% average improvement increases from 19\% with a 2$\times$2 uniform layout to 36\% with a $5{\times}5$ layout. 
% average improvement increases to 36\% with a $5{\times}5$ layout.
However, as the number of tiles grows, the per-tile decode overhead begins to slow queries down.
%  the average improvement for a $7{\times}10$ layout is 28\%. 
\squeezemore{Additionally, variation in performance across videos and queries increases with the number of tiles, as indicated by the widening IQR bars,
% For example, the interquartile range for a $7{\times}10$ layout ranges from 1\% to 58\%. 
demonstrating that the same uniform layout does not work equally well on all videos and queries.}

\subsubsection{Non-uniform tiles} \label{subsec:evalNonUniform}
\tileGranularityPlots

%\fussy
%\textbf{Layout Configuration.}
%evaluates the impact of tile layout objects and tile granularity (illustrated in ~\cref{fig:fineVsCoarseTiles}) on query performance.

The performance of non-uniform layouts depends on the objects queries target and the objects considered when designing the tile layout.
%  which objects are targeted by queries and which objects are considered when designing the tile layout. 
 \Cref{fig:tileGranularityPlot} shows results from different settings. We classify layouts as \textit{same}, \textit{different}, \textit{all}, or \textit{superset}. ``Same'' describes a tile layout around the query object. ``Different'' describes a layout around an object different from the query object (e.g., tiling around people but querying for cars). ``All'' describes tiling around \textit{all} objects detected in the video. Finally, ``superset'' evaluates tiling around the target object and only 1-2 other, frequently occurring objects (e.g., tiling around cars \textit{and} people, as in \cref{subfig:customLayoutCarPed}).
We further classify videos as \textit{sparse}, where the average area occupied by all objects in a frame is ${<}20\%$, or \textit{dense}, where
%  the average area occupied by all objects 
it is ${\geq}20\%$. \Cref{fig:tileGranularityPlot} shows the results. 
\squeeze{The ``different'' and ``superset'' categories only use Visual Road videos and El Fuente scenes that feature multiple object classes; the other videos have a single primary object type.}
% In this experiment, we only use Visual Road videos and El Fuente scenes that feature multiple object classes in the ``different'' and ``superset'' categories; the other videos primarily feature a single type of object. 

%In many cases, users will query for a restricted set of object classes, so we also evaluate the performance of tile layouts around the query object in addition to 1-2 other frequently occurring objects in each video, referred to as

\Cref{fig:tileGranularityPlot} shows that tiling generally improves performance in sparse videos more than dense videos, and tile granularity has the largest impact when objects are dense.
%  achieve larger improvements with tiling than dense videos, and tile granularity has the largest impact on performance when objects are dense.
\Cref{subfig:same} shows that when the tile layout is constructed around the query object, both coarse- and fine-grained tiles significantly improve query performance. 
% The average improvement for fine-grained tiles is 79\% and 51\% for sparse and dense videos, respectively. It is 77\% and 42\% for coarse-grained tiles.
\Cref{subfig:different} shows that tiling around an object type different from the query object hurts performance when objects are dense. This happens when one object is more dense than the others. %, meaning it occupies more area on each frame. 
Querying for the dense object using a layout around the sparse object requires TASM to decode most of the tiles because the dense object occupies much of each frame. Querying for a sparse object using a layout around the dense object also requires most of the frame to be decoded because tiles around dense objects tend to be large. TASM avoids creating these ineffective layouts around dense objects using the decision rule from \cref{sec:knownWorkload}, which we evaluate in \cref{sec:notTilingEval}.
% Improvement in sparse videos is reduced to an average of 41\% and 36\% for fine- and coarse-grained tiles, respectively.
% when fine-grained tiles are used and 36\% with coarse-grained tiles. 
% Query times in spa improve in sparse videos despite using a layout designed for a different object; 
Improvement in sparse videos is reduced, but still positive;
although the query object may intersect multiple tiles, TASM still performs less work if the tiles are small.
% Performance at both granularities suffers when the object types do not appear on the same frames.

\Cref{subfig:all} shows that tiling around all objects is effective only when objects are sparse.
% ; the median improvement for sparse videos is 68\% and 50\% for fine- and coarse-grained tiles, respectively. 
% However, when objects are dense,
% 68\% for fine-grained tiles, and 50\% for coarse-grained tiles. However, when objects are dense, 
% and occupy much of each frame,
% tiling around all objects is \textit{not} effective. 
When objects are dense, median improvement 
% for fine-grained tiles is 21\%, but it 
is 1\% \textit{worse} for coarse-grained tiles.
\Cref{subfig:superset} shows that the ``superset'' strategy performs similarly to ``all''; \squeeze{considering only two or three types of objects rather than all objects when designing layouts achieves small performance gains.}

These results show that tiling around 
% an object (or objects) 
anything
other than the query object slows queries down compared to tiling around the query object. However, fine-grained tiles can still lead to moderate performance improvements in these cases because they are smaller, so fewer non-object pixels must be decoded. % even when the tiles are not designed around the query object.

\tileLayoutDurationVsQueryTimePlot

\textbf{Sequence of tiles (SOT) duration.} Here we evaluate the impact of SOT duration (the number of frames with the same layout) on the performance of non-uniform tile layouts. SOT duration affects the sizes of both tiles and the video. Layout changes must happen at GOP boundaries, so short SOTs require short GOPs and lead to larger storage sizes (see \cref{sec:videoencoding}).
%As described in \cref{sec:videoencoding}, shorter GOPs lead to larger storage sizes.

\sloppy
\Cref{fig:tileLayoutDurationPlot} shows the effect of SOT duration on query performance and storage size. The tiled videos are encoded with a GOP length equal to the SOT duration. We compare query performance and storage size to an untiled video encoded with one-second GOPs (the default GOP duration in most video encoders). Shorter SOT durations lead to larger improvements in query performance because the tiles are smaller and contain fewer non-object pixels.
% The average improvement in query performance decreases from 53\% for one-second SOT durations to 36\% for five-second SOT durations. 
However, shorter SOTs lead to larger storage costs 
% compared to videos encoded with longer SOTs 
because there are more keyframes. 
% Videos encoded with one-second SOTs are an average of 5\% smaller than the original video, while videos encoded with five-second SOTs are on average 15\% smaller.
% smaller than the original.
Note that we see a small improvement in the size of the tiled video with one-second SOTs compared to the original video (also encoded with one-second GOPs); this is due to video encoders being inherently lossy and having the ability to exploit additional compression opportunities during recompression.
% None of the evaluated videos contain prolonged static scenes. If that were the case, we expect SOT duration would not have a large performance impact.
% None of the evaluated videos contain prolonged static scenes, which we expect would decrease the performance impact of SOT duration.
These results demonstrate that setting SOT duration to GOP length is optimal since it leads to the best performance without storage overhead.

\subsubsection{Not tiling} \label{sec:notTilingEval}
\tilingHelpfulnessCutoff

There are videos where tiling is an ineffective strategy to improve query performance. To identify cases where tiling should not be used, we evaluate the effectiveness of a decision rule based on the number of pixels decoded with a given layout. \Cref{fig:tilingCutoff} plots the improvement in query time against the ratio of pixels decoded with a non-uniform layout compared to the untiled video (i.e., $P(v, q, L) / P(v, q, \omega)$) for various videos and query objects. 
\revisioncolor{The figure includes a sampling of diverse layouts, both optimal and suboptimal. The ``same'' category includes the greatest variety of layouts measured, including suboptimal layouts.}
\revisioncolor{While many points overlap, the key observation is that queries for sparse objects primarily lie in the top-left quadrant. This aligns with the expected improvements based on the cost function described in \cref{sec:notation}.}
Using a threshold of not tiling when $P(v, q, L) / P(v, q, \omega) {>} 0.8$ captures nearly all tile layouts that slow queries down (i.e., the improvement is negative). A small number of videos achieve minor performance improvements (${<}20\%$) above this threshold (the upper-right quadrant).

\subsection{Incremental tiling} \label[section]{sec:workloads}

We next evaluate strategies for incremental tiling over various \revision{R3D3}{subframe selection} workloads, which we construct to represent possible query patterns over videos. The baseline strategies are not tiling the video (``Not tiled'') and tiling around all detected objects before queries are processed (``All objects''). 
We compare against two incremental strategies. %these baselines against two incremental strategies. 
``Incremental, more''
%  re-tiles GOPs after observing a query for a new object type (``Incremental, more''). It 
 re-tiles each GOP with a non-uniform, fine-grained layout around all object classes that have been queried so far. 
 For example, if a GOP were queried for cars and then people, TASM would first tile around cars and then re-tile around cars and people.
%  For example, if a GOP is queried for cars, TASM would tile that GOP around cars. If the next query is for people, TASM would re-tile it around cars and people. 
 \squeeze{
 Finally, we evaluate the regret-based approach from \cref{sec:unknownWorkloadUnknownObjects} (``Incremental, regret''). In this strategy, TASM tracks alternative layouts based on the objects queried so far, and re-tiles GOPs once the regret for a layout exceeds the estimated re-encoding cost if the layout is not expected to hurt performance. %TASM estimates the layout will not hurt performance.
 }

TASM estimates the layout will hurt performance if, for any query, $P(s_i, q_i, L) {\geq} \alpha {\cdot} P(s_i, q_i, \omega)$, where $\alpha{=}0.8$ (see \cref{sec:knownWorkload}). TASM estimates the regret using the cost function described in \cref{sec:notation}. Similarly, the re-encoding cost is estimated using a linear model based on the number of pixels being encoded. It was fit based on the time to encode videos with the various layouts used in the microbenchmarks.

% As we are focused on the operations at the storage level, we measure the cumulative time to read video from disk, decode it, and re-tile it with new layouts.
As we are focused on the operations at the storage level, we measure the cumulative time to read video from disk and decode it to answer each query, and re-tile it with new layouts as needed.
% All plots show the cumulative decode \magda{can we say that we show the query time and re-tiling time? And the query time is the read-from-disk and decode time?} and re-tiling time normalized \magda{How is the normalization done? Figure 12 results do not seem to be normalized but absolute.} to the time to execute the queries on the untiled video. 
% Query time includes the time to read the video from disk and decode it.
% Re-tiling time is the time spent re-encoding the video with new layouts.
% The decode time refers to the time spent decoding the pixels requested by each query. The re-tiling time is the time spent re-encoding the video with new layouts.
The time to initially tile the video around all objects is included with the first query for the ``all objects'' strategy. 
We normalize each query's cost to the time to execute that query on the untiled video,
%  to include results for both 2K and 4K videos on the same y-scale
so each query with the ``not tiled'' strategy has a cost of 1.
\squeeze{The lines in \cref{fig:visualroadWorkloads} show the median over all videos the workload was evaluated on.
    % For each plot in \cref{fig:visualroadWorkloads}, the line shows the median over all videos the workload was evaluated on.
%  \Cref{table:workloadResults} contains the IQR. 
%  The first four workloads are evaluated on Visual Road videos where tiling around all objects performs well because objects are sparse. 
%  We evaluate the last two workloads on videos and scenes with dense objects, so tiling around all objects performs poorly.}
% We evaluate the last two workloads on videos and scenes with dense objects.}
We evaluate the first four workloads on Visual Road videos, which have sparse objects, and the last two on videos and scenes with dense objects.}

% \magda{For Figure 12, the y-axes are missing units.}

\workloadFigures
% \workloadResultsTable

% \magda{These results are ok, but they are not presented in an exciting fashion. We should start with the big, impressive results: ``As the figure shows, regret consistently performs best across all methods, except for Workload 1.'' And then go into other results.}
As \cref{fig:visualroadWorkloads} shows, the regret-based approach consistently performs best across all evaluated methods, except for Workload 1. TASM's regret-based approach was designed for more dynamic workloads than Workload 1 where the same query is evaluated across the entire video. For this type of workload, running object detection and tiling up front is a reasonable strategy because all of the results will be used.

We now drill down in the results of each workload.  Queries in Workload 1 target a single object class across the entire video. The workload consists of 100 one-minute queries for cars uniformly distributed over each Visual Road video.
As shown in \cref{subfig:workload1} and discussed above, pre-tiling around all objects performs well when queries 
target % are uniformly distributed over 
the entire video. Incrementally tiling without regret also performs well because all queries target the same object, so SOTs are re-tiled to a layout that speeds up future queries. \squeezemore{The regret-based approach performs poorly over a small number of queries because TASM must observe multiple queries over the same SOT before enough regret accumulates to re-tile. This requires many total queries to be executed when they are uniformly distributed over the entire video.}

% We next evaluate Workload 2 which contains queries for both cars and people. Again, it consists of 60 one-minute queries over the Visual Road videos. Each query had a 50\% chance of being for cars or people. The start frames of each query were picked according to a Zipfian distribution, so frames at the start of the video were more likely to be queried. As shown in \cref{subfig:workload2}, incrementally tiling with regret performs very well. TASM creates layouts around cars and people without wasting time creating intermediate layouts around a single object, like the ``incremental, more'' strategy does. Additionally, the regret-based approach is less likely to create layouts around slices at the end of the video that are infrequently queried.

We next evaluate Workload 2, which examines the performance when queries are restricted to a subset of the video. Workload 2 consists of 100 one-minute queries over the first 25\% of each Visual Road video. Each query has a 50\% chance of being for cars or people. 
% The queries are restricted to frames in the first 25\% of the video. 
As shown in \cref{subfig:workload2}, both incremental strategies perform similarly well. Both outperform pre-tiling the entire video around all objects, which is wasteful when only a small portion of the video is ever queried.

Workload 3 measures the performance when queries are biased towards one section of a video
%  and while the queries generally target the same object classes,
and particular object types. %  and a minority of queries target a less-common object class. 
It consists of 100 queries over the Visual Road videos, where each query has
a 47.5\% chance of being for cars or people, %a 47.5\% chance of being for cars, a 47.5\% chance of being for people
and a 5\% chance of being for traffic lights. 
We exclude % This workload excludes 
one 4K video that did not contain a traffic light. 
The start frame of each query is picked following a Zipfian distribution, so queries are more likely to target the beginning of the video. 
% The queries follow a Zipfian distribution, so queries 
% tend towards % are more likely to target 
% the beginning of the video. \enhao{I didn't understand this sentence for the first time. I think the other version (which you comment out) is more clear.}
As shown in \cref{subfig:workload3}, the regret-based approach performs better than incrementally tiling around more objects because it spends less time re-tiling sections of the video with tile layouts designed around the rarely-queried object.

Workload 4 measures performance when queries target different objects over time. It consists of 200 one-minute queries
following a Zipfian distribution 
over the Visual Road videos.
The middle third of the queries target people, and the rest target cars. % where the first third of queries are for cars, the middle third for people, and the final third for cars. 
% The %start frames of the 
% queries follow a Zipfian distribution. 
As shown in \cref{subfig:workload4}, the incremental, regret-based approach performs well and does not exhibit large jumps in decode and re-tiling time when the query object changes.

% Workload 6 measures the performance of the strategies when different objects are queried for in different sections of the video. The workload consists of 60 one-minute queries over the Visual Road videos where queries uniformly distributed over the first 50\% of each video. Queries that started in the first 25\% of the video were for cars, and queries that started in the second 25\% were for people. As shown in \cref{subfig:workload6}, the ``incremental, more'' strategy is able to quickly create layouts around the objects that are targeted in each section of the video because each slice sees queries for the same object over time. The incremental, regret-based approach eventually creates useful layouts, but it takes longer because it observes more queries before re-tiling a section.

Workload 5 measures performance when tiling is not effective. % an effective strategy. 
It is evaluated on select videos from the Xiph, Netflix public dataset, and scenes from the El Fuente video that contain diverse scenes with many types of objects (e.g., markets with people, cars, and food).
% (e.g., markets filled with people, parked cars, and different kinds of food). 
The queries are uniformly distributed, and each randomly targets one of the video's primary objects within one-second. As shown in \cref{subfig:workload5}, only the regret-based approach keeps costs similar to not tiling. ``All objects'' performs poorly because objects are dense in these scenes. ``Incremental, more'' performs poorly because it spends time re-tiling with layouts that perform similarly to the untiled video.

Finally, Workload 6 measures performance when tiling around the query object is beneficial, but tiling around all objects is not. 
It is evaluated on select videos from the Netflix public dataset and scenes from the full El Fuente video 
that fit this criteria.
% where tiling around one object performs well, but tiling around all objects performs poorly. 
The queries are uniformly distributed, and each targets the same object class over one second. As shown in \cref{subfig:workload6}, both incremental strategies eventually achieve layouts that perform better than not tiling. 
\squeeze{``All objects'' performs poorly because objects in these videos are dense.}

\subsection{\revision{R1W3, R1D1}{Macrobenchmark}}\label{sec:macrobenchmarkEval}
\visualRoadBenchmarkFigure

% \magda{We need to say somewhere what are those 6 videos???}
% \magda{Are the spikes when object detection IS performed or when it DOES NOT NEED to be performed?}

\revisioncolor{
Beyond the decoding benchmarks, we also evaluate TASM's performance on an end-to-end workload from the Visual Road benchmark~\cite{DBLP:conf/sigmod/HaynesMBCC19}, specifically Q7.
Each query in the workload specifies a temporal range and a set of object classes. The following tasks are executed per-query: (i) detect objects if not previously done on the specified temporal range, (ii) draw boxes around the specified object classes, and (iii) encode the modified frames. The original Visual Road query involves masking the background pixels, but we omit that step
to demonstrate TASM's benefits when users want to view full frames. We compare the performance of executing this query on untiled frames to TASM with
incremental, regret-based tiling.
We detect objects by running YOLOv3~\cite{DBLP:journals/corr/abs-1804-02767} every three frames.
TASM adds bounding boxes by decoding only the tiles that contain the requested objects, drawing the boxes, then re-encoding these tiles. TASM outputs the full frame by homomorphically stitching the modified tiles that contain the object with the original tiles that do not contain the object.
}

\revisioncolor{
We execute 100 one-minute queries over the Visual Road videos, using a Zipfian distribution over time-ranges. Each query is randomly for cars or people. 
\Cref{fig:visualRoadBenchmarkFigure} shows the median speedup achieved with TASM compared to the untiled video over six orderings of the queries.
% We execute this workload with different query orders to smooth order-dependent spikes in the speedup that appear when object detection is performed.
TASM reduces the total workload runtime by 12{-}39\% across the videos.
Object detection contributes significantly to the total runtime and LightDB does not use a pre-filtering step to accelerate this operation.
If we examine one instance of the workload where the last 20 queries no longer need to perform object detection and execute after TASM has found good layouts, the median improvement for these queries ranges from $23\%$ to $66\%$ across the videos.
While these queries request the full frame, TASM accelerates them by processing just the relevant regions of the frame, which allows it to decode and encode less data.
}

\subsection{Object detection acceleration}\label{sec:objectDetectionEval}

\objDetectionResultsCombined

% \magda{Figure 14: Caption is too short and thus unclear. We should make sure FPS is expanded in the text or in the caption. In the text, we need to explain what each column means including the Python vs C++ columns.}
% \magda{Table 3. Caption is too short and thus cunclear. Can put both Fig 14 and Table III into one figure with one caption. Also, units in the table are missing.}
% \magda{Importantly: The reader may not have read the BlazeIt paper, so they will have really no idea what we are talking about here. We need to describe what is happening more clearly for a reader who doesn't know BlazeIt.}
We now evaluate TASM's ability to accelerate the full scan phase of object detection queries, as described in \cref{sec:introduction}.
One system that uses specialized models during the full scan phase %to rapidly compute queries 
is BlazeIt~\cite{DBLP:journals/pvldb/KangBZ19}.
For example, it uses a specialized counting model to compute aggregates.
% For example, it uses a model that predicts the number of objects of a particular type in each frame when evaluating aggregate queries.
We evaluate TASM's ability to accelerate this phase using BlazeIt's counting model as a representative fast model.
This
% specialized counting 
model runs at over 1K frames per second (fps), while preprocessing the frames % and preparing them as input to the model 
runs below 300 fps. TASM reduces the preprocessing bottleneck and achieves up to a $2{\times}$ speedup while maintaining the model's accuracy.

The preprocessing phase includes reading video from disk, decoding and resizing frames, normalizing pixel values, and transforming the pixel format.
BlazeIt implements this using Python, OpenCV~\cite{opencv}, and Numpy~\cite{2020NumPy-Array} (``Python'' in \cref{fig:preprocessingTput}).
We reimplemented this using C++, NVDECODE~\cite{nvenc}, and Intel IPP~\cite{ipp} to fairly compare against TASM (``C++'').
We evaluate on three days of BlazeIt's \texttt{\small{grand-canal}} video dataset.

We compare against using semantic predicate pushdown with ROI layouts generated by TASM.
We first use \revision{R1D4}{MOG2}-based background segmentation implemented in OpenCV~\cite{opencv} to detect foreground ROIs 
\revision{R1D4}{on the first frame of each GOP. This is a throughput that recent mobile devices are known to operate above~\cite{fastestBackgroundSubtraction}, and therefore it would be possible for this step to be offloaded to a compute-enabled camera as discussed in \cref{sec:ROI}.}
% \brandon{Yeah, it's too abrupt and needs a bit of smoothing.  The trick here is to talk about modeling an edge device without actually running the experiment on.  Lots of `We elected to run MOG2 at X FPS, which is a throughput that recent mobile devices are known to operate at [cite].'}
We use TASM to create fine-grained tiles (``Fine tiles'') and coarse-grained tiles (``Coarse tiles'') around the foreground regions.
\squeeze{We also compare against a tile layout created around a manually-specified ROI capturing the canal in the lower-left
portion of each frame (``ROI'').}

% We created both coarse and fine-grained tiles around foreground regions detected by KNN-based background segmentation implemented in OpenCV~\cite{opencv}, as well as a layout around a manually specified ROI that captures the lower-left portion of each frame.
% We first measure the preprocessing \magda{Confusing what we mean by preprocessing throughput. We measure the preprocessing phase but when we talk about preprocessing it sounds like the preprocessing phase or our approach. Do you mean the time to read from disk and decode?} throughput as it is the bottleneck; the specialized model can achieve a throughput of over 1K fps.
\Cref{fig:preprocessingTput} shows the preprocessing throughput when operating on entire frames compared to just the tiles that contain ROIs.
Operating on tiles improves throughput by up to $2{\times}$ %achieves up to $2\times$ improvement in throughput 
and therefore reduces the bottleneck for performing inference with the specialized model.
\squeeze{
We next %measure the accuracy of the specialized model
% when operating over tiles to verify that the improved preprocessing performance
verify that using tiles rather than full frames
does not negatively impact the model's accuracy.
We use the same %counting 
model architecture for tiled inputs. However, rather than training and inferring using full frames, we use a single tile from each frame that contains all ROIs.  % hence why we evaluate accuracy on the coarse-grained layout. % layout rather than the fine-grained layout.
For each strategy we train BlazeIt's counting model on the first 150K frames or tiles from the first day of video. We evaluate this model on 150K frames or tiles from each day (using a different set of frames for the first day).
As shown in \cref{table:Accuracy}, models trained and evaluated on tiles show similar accuracy to full frame training \revision{R2D2}{within each day}.
}
%!TEX root = paper.tex

\section{Related work}

% Systems that use specialized techniques for fast inference over videos.
% TASM focuses on optimizing query execution at the storage layer, while many recent VDBMSs accelerate queries using other techniques.  
As mentioned in \cref{sec:introduction}, many systems optimize extracting semantic content from videos.
BlazeIt~\cite{DBLP:journals/pvldb/KangBZ19} and NoScope~\cite{DBLP:journals/pvldb/KangEABZ17} apply specialized NNs that run faster than general models.
Other systems filter frames before applying expensive models: probabilistic predicates~\cite{DBLP:conf/sigmod/LuCKC18} and ExSample~\cite{DBLP:journals/corr/abs-2005-09141} use statistical techniques, MIRIS~\cite{DBLP:conf/sigmod/BastaniHBGABCKM20} uses sampling, and SVQ~\cite{DBLP:conf/sigmod/XarchakosK19} and \textit{IC} and \textit{OD} filters~\cite{DBLP:conf/icde/KoudasLX20} use deep learning filters.
These systems and techniques can use TASM to run models on specific ROIs to reduce their preprocessing overhead.
% Additionally, information about the semantic content of videos generated by these systems could be incorporated into TASM's semantic index to inform tiling decisions.
Focus~\cite{Focus:osdi/HsiehABVBPGM18} shifts some processing to ingest-time.
Systems such as LightDB~\cite{LightDB:pvldb/HaynesMABCC18}, Optasia~\cite{DBLP:conf/cloud/LuCK16}, and Scanner~\cite{Scanner:tog/PomsCHF18} accelerate queries through parallelization and deduplication of work, while VideoEdge~\cite{DBLP:conf/edge/HungABGYBP18} distributes processing over clusters. These general VDBMSs could incorporate TASM to further accelerate performance.
\squeeze{Panorama~\cite{Panorama:pvldb/ZhangK19} and Rekall~\cite{Rekall:corr/abs-1910-02993} expand the set of queries that can be executed over videos, which is orthogonal to video storage.}

% Systems that support general query processing over videos.
% Scanner, Optasia, LightDB

% Another class of VDBMSs supports general query processing over video and could incorporate a TASM to further accelerate query processing.
% accelerates general queries over video by parallelizing them over many machines.  optimizes queries over multiple cameras through parallelization in addition to deduplicating work within a query.  accelerates queries over virtual and augmented reality videos through parallelization and techniques that exploit the structure of encoded videos.

% Systems that optimize the storage of videos.
Other systems also target storage-level optimizations.
VStore~\cite{VStore:eurosys/XuBL19} modifies encoding parameters to accelerate processing while maintaining accuracy.
% It improves performance by reducing the quality of the video, while TASM attempts to maintain video quality.
Smol~\cite{DBLP:journals/corr/abs-2007-13005} jointly optimizes video resolution and NN architectures to achieve high accuracy while accelerating preprocessing, but, like VStore, only considers reducing the resolution of videos while TASM maintains video quality.
% Additionally, VStore must profile all downstream operators to determine the encoding parameters, while TASM can work incrementally as queries are processed.
Vignette~\cite{DBLP:conf/cloud/MazumdarHBCCO19} uses tiles for perception-based compression but only considers uniform layouts.

% Systems that support general queries over videos.

%  Panorama~\cite{Panorama:pvldb/ZhangK19} introduces a multi-task CNN architecture that can generalize to unbounded vocabularies, and Rekall~\cite{Rekall:corr/abs-1910-02993} introduces a framework to specify events by composing the outputs of many models.

% Related work in relational databases for incremental indexing.
TASM's incremental tiling approach is inspired by database cracking~\cite{DBLP:conf/cidr/IdreosKM07, DBLP:journals/pvldb/HalimIKY12}, which incrementally reorganizes the data processed by each query, and online indexing~\cite{DBLP:conf/icde/BrunoC07} which creates and modifies indices as queries are processed. Regret has also been used to design an economic model for self-tuning indices in a shared cloud database~\cite{DBLP:conf/icde/DashKA09}. \squeeze{TASM extends these relational storage techniques to provide efficient access to video data.}

% Moving compute to edge
% TASM proposes utilizing the improving compute capacity of edge devices to generate the semantic index and tile layouts on-camera. Achieving real-time object detection on resource-limited edge devices is an active area of research. Systems like MARLIN~\cite{DBLP:conf/sensys/Apicharttrisorn19} speed up object detection on-device by combining expensive object detection with cheap object tracking methods. Alternative approaches distribute computation between the edge device and cloud, like MCDNN~\cite{DBLP:conf/mobisys/HanSPAWK16}, Rocket~\cite{DBLP:journals/computer/Ananthanarayanan17}, and  DeepDecision~\cite{DBLP:conf/infocom/RanCZLC18}, or use smaller models to provide faster inference on the edge, such as MobileNets~\cite{DBLP:journals/corr/HowardZCKWWAA17} and YOLOv3-tiny~\cite{DBLP:journals/corr/abs-1804-02767}.
% Frameworks have also been proposed to accelerate inference on mobile hardware through efficient allocation of GPU and CPU resources~\cite{DBLP:conf/mobisys/LocLB17, DBLP:conf/mobicom/FangZ018}.
% combine object detection models with fast object tracking methods on frames that are captured between calls to the detector.

% The observation that it can be useful to retrieve spatial subsets of videos has come up in other application domains.
Other application domains have observed the usefulness of retrieving spatial subsets of videos.
The MPEG DASH SRD standard~\cite{DBLP:conf/mmsys/NiamutTDCDL16} is motivated by a similar observation that video streaming clients occasionally request a spatial subset of videos. While it specifies a model to support streaming spatial subsets, it does not specify \textit{how} to efficiently partition videos into tiles.

\section{Conclusion}
%This paper introduces %the design, implementation, and evaluation of
We presented
TASM, a tile-based storage manager that accelerates \revision{R3D3}{subframe selection} and object detection queries by targeting spatial frame subsets.
%We proposed strategies that allow 
TASM incrementally tiles sections of the video as queries execute, leading to improved performance (up to 94\%).
%  and objects are detected
%and demonstrate that 
\squeeze{
We also showed how TASM alleviates bottlenecks 
%during a full scan 
%of object detection queries 
by only reading areas likely to contain objects.
}

\textbf{Acknowledgments.}
This work was supported in part by NSF award CCF-1703051, an award from the University of Washington Reality Lab, a gift from Intel, and DARPA through RC Center grant GI18518. A. Mazumdar was also supported in part by a CoMotion Commercialization Fellows grant.

\bibliographystyle{IEEEtran}
\bibliography{IEEEabrv,header,others,self}

\begin{thebibliography}{10}
\providecommand{\url}[1]{#1}
\csname url@samestyle\endcsname
\providecommand{\newblock}{\relax}
\providecommand{\bibinfo}[2]{#2}
\providecommand{\BIBentrySTDinterwordspacing}{\spaceskip=0pt\relax}
\providecommand{\BIBentryALTinterwordstretchfactor}{4}
\providecommand{\BIBentryALTinterwordspacing}{\spaceskip=\fontdimen2\font plus
\BIBentryALTinterwordstretchfactor\fontdimen3\font minus
  \fontdimen4\font\relax}
\providecommand{\BIBforeignlanguage}[2]{{%
\expandafter\ifx\csname l@#1\endcsname\relax
\typeout{** WARNING: IEEEtran.bst: No hyphenation pattern has been}%
\typeout{** loaded for the language `#1'. Using the pattern for}%
\typeout{** the default language instead.}%
\else
\language=\csname l@#1\endcsname
\fi
#2}}
\providecommand{\BIBdecl}{\relax}
\BIBdecl

\bibitem{DBLP:conf/cloud/LuCK16}
Y.~Lu \emph{et~al.}, ``Optasia: {A} relational platform for efficient
  large-scale video analytics,'' in \emph{{SoCC}}, 2016, pp. 57--70.

\bibitem{DBLP:conf/nsdi/ZhangABPBF17}
H.~Zhang \emph{et~al.}, ``Live video analytics at scale with approximation and
  delay-tolerance,'' in \emph{{NSDI}}, 2017, pp. 377--392.

\bibitem{DBLP:journals/pvldb/KangBZ19}
D.~Kang \emph{et~al.}, ``{BlazeIt}: Optimizing declarative aggregation and
  limit queries for neural network-based video analytics,'' \emph{PVLDB},
  vol.~13, 2019.

\bibitem{DBLP:conf/mobicom/WangCC16}
X.~Wang \emph{et~al.}, ``Skyeyes: adaptive video streaming from uavs,'' in
  \emph{HotWireless@MobiCom}, 2016, pp. 2--6.

\bibitem{DBLP:conf/edge/WangFCGBPYS18}
J.~Wang \emph{et~al.}, ``Bandwidth-efficient live video analytics for drones
  via edge computing,'' in \emph{{SEC}}, 2018, pp. 159--173.

\bibitem{Panorama:pvldb/ZhangK19}
Y.~Zhang \emph{et~al.}, ``Panorama: {A} data system for unbounded vocabulary
  querying over video,'' \emph{{PVLDB}}, vol.~13, pp. 477--491, 2019.

\bibitem{DBLP:journals/pvldb/KangEABZ17}
D.~Kang \emph{et~al.}, ``Noscope: Optimizing deep {CNN}-based queries over
  video streams at scale,'' \emph{{PVLDB}}, vol.~10, pp. 1586--1597, 2017.

\bibitem{Scanner:tog/PomsCHF18}
A.~Poms \emph{et~al.}, ``Scanner: efficient video analysis at scale,''
  \emph{{ACM} Trans. Graph.}, vol.~37, pp. 138:1--138:13, 2018.

\bibitem{Focus:osdi/HsiehABVBPGM18}
K.~Hsieh \emph{et~al.}, ``Focus: Querying large video datasets with low latency
  and low cost,'' in \emph{{OSDI}}, 2018, pp. 269--286.

\bibitem{VStore:eurosys/XuBL19}
T.~Xu \emph{et~al.}, ``{VStore}: {A} data store for analytics on large
  videos,'' in \emph{EuroSys}.\hskip 1em plus 0.5em minus 0.4em\relax {ACM},
  2019, pp. 16:1--16:17.

\bibitem{DBLP:conf/sigmod/BastaniHBGABCKM20}
F.~Bastani \emph{et~al.}, ``{MIRIS:} fast object track queries in video,'' in
  \emph{{SIGMOD}}, 2020, pp. 1907--1921.

\bibitem{DBLP:conf/sigmod/LuCKC18}
Y.~Lu \emph{et~al.}, ``Accelerating machine learning inference with
  probabilistic predicates,'' in \emph{{SIGMOD}}.\hskip 1em plus 0.5em minus
  0.4em\relax {ACM}, 2018, pp. 1493--1508.

\bibitem{DBLP:journals/corr/abs-2007-13005}
D.~Kang \emph{et~al.}, ``Jointly optimizing preprocessing and inference for
  {DNN}-based visual analytics,'' \emph{CoRR}, vol. abs/2007.13005, 2020.

\bibitem{waymoDatasetFAQ}
Waymo, ``Waymo open dataset {FAQ},'' \url{waymo.com/open/faq}, 2021.

\bibitem{DBLP:conf/eusipco/AbuTahaSHDVV18}
M.~AbuTaha \emph{et~al.}, ``End-to-end real-time {ROI}-based encryption in
  {HEVC} videos,'' in \emph{{EUSIPCO}}.\hskip 1em plus 0.5em minus 0.4em\relax
  {IEEE}, 2018, pp. 171--175.

\bibitem{DBLP:conf/cidr/IdreosKM07}
S.~Idreos \emph{et~al.}, ``Database cracking,'' in \emph{{CIDR}}, 2007, pp.
  68--78.

\bibitem{DBLP:journals/pvldb/HalimIKY12}
F.~Halim \emph{et~al.}, ``Stochastic database cracking: Towards robust adaptive
  indexing in main-memory column-stores,'' \emph{{PVLDB}}, vol.~5, 2012.

\bibitem{DBLP:conf/icde/BrunoC07}
N.~Bruno \emph{et~al.}, ``An online approach to physical design tuning,'' in
  \emph{{ICDE}}, 2007, pp. 826--835.

\bibitem{DBLP:conf/icde/DashKA09}
D.~Dash \emph{et~al.}, ``An economic model for self-tuned cloud caching,'' in
  \emph{{ICDE}}, 2009, pp. 1687--1693.

\bibitem{H264Specification}
``Advanced video coding for generic audiovisual services,'' Rec. ITU-T H.264
  and ISO/IEC 14496-10, 06 2019.

\bibitem{HEVCSpecification}
``High efficiency video coding,'' Rec. ISO/IEC 23008-2, Nov 2019.

\bibitem{AV1Specification}
P.~de~Rivaz \emph{et~al.}, ``{AV1} bitstream \& decoding process
  specification,'' \emph{The Alliance for Open Media}, p. 182, 2018.

\bibitem{DBLP:journals/tcsv/SullivanOHW12}
G.~J. Sullivan \emph{et~al.}, ``Overview of the high efficiency video coding
  {(HEVC)} standard,'' \emph{TCSVT}, vol.~22, pp. 1649--1668, 2012.

\bibitem{LightDB:pvldb/HaynesMABCC18}
B.~Haynes \emph{et~al.}, ``{LightDB}: {A} {DBMS} for virtual reality video,''
  \emph{{PVLDB}}, vol.~11, pp. 1192--1205, 2018.

\bibitem{DBLP:conf/sigmod/ChaudhuriN98}
S.~Chaudhuri \emph{et~al.}, ``Autoadmin 'what-if' index analysis utility,'' in
  \emph{{SIGMOD}}, 1998, pp. 367--378.

\bibitem{boschCameraTrainer}
``Bosch camera trainer with fw 7.10 and cm 6.20,''
  \url{boschsecurity.com/us/en/solutions/video-systems/video-analytics/technical-documentation-for-video-analytics},
  2019, accessed: 2020-06.

\bibitem{fastestBackgroundSubtraction}
S.~Zeevi, ``{BackgroundSubtractorCNT},''
  \url{https://sagi-z.github.io/BackgroundSubtractorCNT}, 2016, accessed:
  2021-01-21.

\bibitem{DBLP:conf/sigmod/HaynesMBCC19}
B.~Haynes \emph{et~al.}, ``{Visual Road}: {A} video data management
  benchmark,'' in \emph{{SIGMOD}}, 2019, pp. 972--987.

\bibitem{netflixPublicDataset}
Z.~Li \emph{et~al.}, ``Toward a practical perceptual video quality metric,''
  \url{https://netflixtechblog.com/653f208b9652}, 06 2016, accessed:
  2020-06-01.

\bibitem{netflixOpenSourceAssets}
A.~Schuler \emph{et~al.}, ``Engineers making movies ({AKA} open source test
  content),'' \url{netflixtechblog.com/f21363ea3781}, 2018, accessed: 2020-06.

\bibitem{xiphDataset}
``Xiph.org video test media,'' \url{media.xiph.org/video/derf}, 2019.

\bibitem{DBLP:journals/corr/MilanL0RS16}
A.~Milan \emph{et~al.}, ``{MOT16:} {A} benchmark for multi-object tracking,''
  \emph{CoRR}, vol. abs/1603.00831, 2016.

\bibitem{cdvlElFuente}
I.~Katsavounidis, ``Netflix – ``{El Fuente}'' video sequence details and
  scenes,'' \url{cdvl.org/documents/ElFuente_summary.pdf}, accessed: 2020-06.

\bibitem{nvenc}
``Nvidia video codec,'' \url{developer.nvidia.com/nvidia-video-codec-sdk}.

\bibitem{ffmpeg}
F.~Bellard, ``{FFmpeg},'' \url{ffmpeg.org}, 2018.

\bibitem{DBLP:journals/corr/abs-1804-02767}
J.~Redmon \emph{et~al.}, ``Yolov3: An incremental improvement,'' \emph{CoRR},
  vol. abs/1804.02767, 2018.

\bibitem{sqlite}
``Sqlite,'' \url{https://sqlite.org/index.html}, 2020.

\bibitem{DBLP:conf/asplos/LottariniRCKRSW18}
A.~Lottarini \emph{et~al.}, ``vbench: Benchmarking video transcoding in the
  cloud,'' in \emph{{ASPLOS}}, 2018, pp. 797--809.

\bibitem{vranjes2008locally}
M.~Vranjes \emph{et~al.}, ``Locally averaged psnr as a simple objective video
  quality metric,'' in \emph{ELMAR}, vol.~1.\hskip 1em plus 0.5em minus
  0.4em\relax IEEE, 2008, pp. 17--20.

\bibitem{opencv}
OpenCV, ``{Open Source Computer Vision Library},'' \url{opencv.org}, 2018.

\bibitem{2020NumPy-Array}
C.~R. Harris \emph{et~al.}, ``Array programming with {NumPy},'' \emph{Nature},
  2020.

\bibitem{ipp}
Intel, ``Intel integrated performance primitives,'' \url{
  software.intel.com/content/www/us/en/develop/documentation/ipp-dev-reference}.

\bibitem{DBLP:journals/corr/abs-2005-09141}
O.~Moll \emph{et~al.}, ``Exsample: Efficient searches on video repositories
  through adaptive sampling,'' \emph{CoRR}, vol. abs/2005.09141, 2020.

\bibitem{DBLP:conf/sigmod/XarchakosK19}
I.~Xarchakos \emph{et~al.}, ``{SVQ:} streaming video queries,'' in
  \emph{{SIGMOD}}, 2019.

\bibitem{DBLP:conf/icde/KoudasLX20}
N.~Koudas \emph{et~al.}, ``Video monitoring queries,'' in \emph{{ICDE}}, 2020.

\bibitem{DBLP:conf/edge/HungABGYBP18}
C.~Hung \emph{et~al.}, ``{VideoEdge}: Processing camera streams using
  hierarchical clusters,'' in \emph{{SEC}}, 2018, pp. 115--131.

\bibitem{Rekall:corr/abs-1910-02993}
D.~Y. Fu \emph{et~al.}, ``Rekall: Specifying video events using compositions of
  spatiotemporal labels,'' \emph{CoRR}, vol. abs/1910.02993, 2019.

\bibitem{DBLP:conf/cloud/MazumdarHBCCO19}
A.~Mazumdar \emph{et~al.}, ``Perceptual compression for video storage and
  processing systems,'' in \emph{SoCC}, 2019, pp. 179--192.

\bibitem{DBLP:conf/mmsys/NiamutTDCDL16}
O.~A. Niamut \emph{et~al.}, ``{MPEG} {DASH} {SRD:} spatial relationship
  description,'' in \emph{MMSys}, 2016, pp. 5:1--5:8.

\end{thebibliography}

\end{sloppypar}

\end{document}